\documentclass[journal, onecolumn]{IEEEtran}

\usepackage{subfig} 

\usepackage[total={6in,9in},top=1in,left=1.25in,bindingoffset=0in]{geometry}
\usepackage[normalem]{ulem} 
\usepackage{pifont} 
\usepackage[usenames]{color} 
\usepackage{amsmath} 
\usepackage{amsfonts} 
\usepackage{array} 
\usepackage{graphicx} 
\usepackage[english]{babel} 

\usepackage{graphicx} 
\usepackage{amsthm}

\usepackage{pgf}
\usepackage{tikz}
\usepackage{pgfplots}
\usetikzlibrary{arrows,automata}
\usetikzlibrary{positioning}
\usetikzlibrary{backgrounds}
\usetikzlibrary{matrix}
\usetikzlibrary{decorations.pathmorphing}
\usetikzlibrary{fit}

\newtheorem{lemm}{Lemma}
\newtheorem{prop}{Proposition}
\newtheorem{theo}[lemm]{Theorem}
\newtheorem{exmp}{Example}[section]

\newenvironment{demo}{\noindent {\indent{\em Proof: \ }}\ }{\qed}

\title{A realistic distributed storage system: the rack model.}
\author{ \vspace{0.25cm}Bernat Gast\'on, Jaume Pujol, and Merc\`e Villanueva\\
\small Department of Information and Communications Engineering \\
\small Universitat Aut\`onoma de Barcelona \\
\small Cerdanyola del Vall\`{e}s (Barcelona), Spain \\
\small \texttt{\{Bernat.Gaston $|$ Jaume.Pujol $|$
Merce.Villanueva \}@uab.cat} \\
}
\begin{document}
\maketitle

\begin{abstract}
In a realistic distributed storage environment, storage nodes are usually
placed in racks, a metallic support designed to accommodate electronic
equipment. It is known that the communication (bandwidth) cost between nodes
which are in the same rack is much lower than between nodes which are in
different racks.

In this paper, a new model, where the storage nodes are placed in two racks,
is proposed and analyzed. Moreover, the two-rack model is generalized to any
number of racks. In this model, the storage nodes have different
repair costs depending on the rack
where they are placed. A threshold function, which minimizes the amount of
stored data per node and the bandwidth needed to regenerate a failed node, is
shown. This threshold function generalizes the ones given for previous
distributed storage models. The tradeoff curve obtained from this threshold
function is compared with the ones obtained from the previous models, and it is
shown that this new model outperforms the previous ones in terms of repair cost.
\end{abstract}

\section{Introduction}
\label{Intro}

In a distributed storage environment, where the data is
placed in nodes connected through a network, it is likely that one of these
nodes fails. It is known that the use of erasure coding improves the fault
tolerance and minimizes the amount of stored data per node \cite{Rod01}, \cite{We01}.
Moreover, the use of regenerating codes  not only makes the most of the erasure
coding improvements, but also minimizes the bandwidth needed to regenerate
a failed node \cite{Di01}.

In realistic distributed storage environments, for example a storage cloud, the
data is placed in storage devices which are connected through a network. These
storage devices are usually organized in a rack, a metallic support designed to
accommodate electronic equipment. The communication (bandwidth) cost between
nodes which are in the same rack is much lower than between nodes which are in
different racks. In fact, in \cite{An01} it is said that reading from a local
disk is nearly as efficient as reading from the disk of another node in the same
rack.

In \cite{Di01}, an optimal tradeoff given by a threshold function between the
amount of stored data per node
and the bandwidth needed to regenerate a failed node (repair bandwidth)
in a distributed storage environment was claimed. This tradeoff was
proved by using the mincut on information flow graphs, and it can be represented as a
curve, where the two extremal points of the curve are called the Minimum Storage
Regenerating (MSR) point and the Minimum Bandwidth Regenerating (MBR) point.

In \cite{Ak01}, another model, where there is a static classification of
the storage nodes in two sets: one with the ``cheap bandwidth'' nodes, and another one
with the ``expensive bandwidth'' nodes, was presented and analyzed. This classification
of the nodes is not based on racks, because the nodes in the expensive set are
always expensive in terms of the cost of sending data to a newcomer, regardless
of the specific newcomer. A description of this model is included in Subsection
\ref{subsec:02}.
There are other models, usually called non-homogeneous, which are based on one
or more nodes being able to store different amounts of data. Examples of these
models are presented in \cite{Yu01} and \cite{Va01}.

This paper is organized as follows. In Section \ref{sec:1}, we review previous
distributed storage models in order to present the new model in next section.
In Section \ref{sec:2}, we start by describing this new model where the storage nodes are placed in two
racks. We also provide a general threshold function, and we describe the
extremal Minimum Storage and Minimum Bandwidth Regenerating points. In Section
\ref{sec:3}, we generalize the two-rack model to any number of racks. In Section
\ref{sec:4}, we analyze the results obtained from this new model by comparing
them with the previous models. Finally, in Section \ref{sec:5}, we expose the
conclusions of this study.

\section{Previous models}
\label{sec:1}

In this section, we describe the previous distributed storage models: the
basic model and the static cost model introduced in \cite{Di01} and \cite{Ak01}, respectively.
\subsection{Basic model}
\label{subsec:01}

In \cite{Di01}, \textit{Dimakis et al.} introduced a first distributed storage
model, where each storage node has the same repair bandwidth.
Moreover, the fundamental tradeoff between the amount of stored data per node
and the repair bandwidth was given from analyzing the mincut of an information
flow graph.

Let $C$ be a $[n,k,d]$ regenerating code composed by $n$ storage nodes, each
one storing $\alpha$ data units, and such that any $k$ of these $n$ storage
nodes contain enough information to recover the file. In order to be able to
recover a file of size $M$, it is necessary that $\alpha k \ge M$.
When one node fails, $d$ of the remaining $n-1$ storage nodes send $\beta$ data
units to the new node which replaces the failed one. The new node is called
newcomer, and the set of $d$  nodes sending data to the newcomer are called helper
nodes. The total amount of bandwidth used per node regeneration is $\gamma= d
\beta$.

\begin{figure}
\centering
\begin{tikzpicture}[shorten >=1pt,->]
  \tikzstyle{vertix}=[circle,fill=black!25,minimum size=18pt,inner sep=0pt,
node distance = 0.8cm,font=\tiny]
  \tikzstyle{invi}=[circle]
  \tikzstyle{background}=[rectangle, fill=gray!10, inner sep=0.2cm,rounded
corners=5mm]

  \node[vertix] (s) {$S$};

  \node[vertix, right=1cm of s] (vin_2)  {$v_{in}^2$};
  \node[vertix, below of=vin_2] (vin_3)  {$v_{in}^3$};
  \node[vertix, above of=vin_2] (vin_1)  {$v_{in}^1$};
  \node[vertix, below of=vin_3] (vin_4)  {$v_{in}^4$};

 \foreach \to in {1,2}
   {\path (s) edge[bend left=20,font=\tiny] node[anchor=south,above]{$\infty$}
(vin_\to);}
 \foreach \to in {3,4}
   {\path (s) edge[bend right=20,font=\tiny]  node[anchor=south,above]{$\infty$}
 (vin_\to);}

  \node[vertix, right of=vin_1] (vout_1)  {$v_{out}^1$};
  \node[vertix, right of=vin_2] (vout_2)  {$v_{out}^2$};
  \node[vertix, right of=vin_3] (vout_3)  {$v_{out}^3$};
  \node[vertix, right of=vin_4] (vout_4)  {$v_{out}^4$};

  \node[vertix, right of=vout_4, node distance = 2cm] (vin_5)
{\scriptsize{$v_{in}^5$}};
  \node[vertix, right of=vin_5] (vout_5)
{\scriptsize{$v_{out}^5$}};

  \node[vertix, right of=vout_1, above of=vin_5, node distance = 2cm] (vin_6)
{\scriptsize{$v_{in}^6$}};
  \node[vertix, right of=vin_6] (vout_6)
{\scriptsize{$v_{out}^6$}};

  \node[vertix, right of=vout_6, node distance = 1.5cm] (DC)
{DC};

   \path (vout_5) edge[bend right=20,font=\tiny]
node[anchor=south,above]{$\infty$}
 (DC);
   \path (vout_6) edge[bend left=10,font=\tiny]
node[anchor=south,above]{$\infty$}
 (DC);

  \path[->, bend left=15,font=\tiny] (vout_2) edge
node[anchor=south,above]{$\beta$} (vin_5);
  \path[->, bend left=10,font=\tiny] (vout_3) edge
node[anchor=south,above]{$\beta$}(vin_5);
  \path[->, bend left=20,font=\tiny] (vout_1) edge
node[anchor=south,above]{$\beta$}(vin_5);

  \path[->, bend left=15,font=\tiny] (vout_1) edge
node[anchor=south,above]{$\beta$} (vin_6);
  \path[->, bend left=10,font=\tiny] (vout_2) edge
node[anchor=south,above]{$\beta$}(vin_6);
  \path[->, bend left=10,font=\tiny] (vout_5) edge
node[anchor=south,above]{$\beta$}(vin_6);

 \foreach \from/\to in {1,2,3,4,5,6}
  { \path[->,font=\tiny] (vin_\from) edge node[anchor=south] {$\alpha$}
(vout_\to); }

  \draw[-,color=red,thick] (1.2,-0.5) -- (2.7,-1.2);
  \draw[-,color=red,thick] (1.2,-1.2) -- (2.7,-0.5);

  \draw[-,color=red,thick] (1.2,-1.2) -- (2.7,-1.9);
  \draw[-,color=red,thick] (1.2,-1.9) -- (2.7,-1.2);
\end{tikzpicture}

\caption{Information flow graph corresponding to a $[4,2,3]$ regenerating
code.}
\label{Fig:01}
\end{figure}
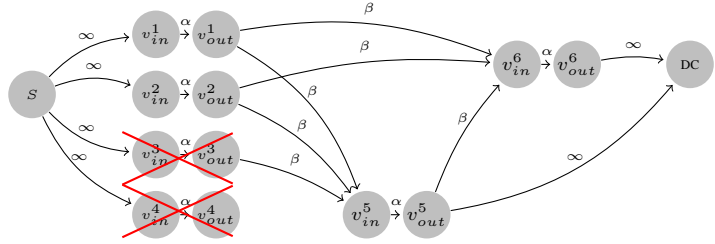

Let $s_i$, where $i = 1,\ldots, \infty$, be the $i$-th storage node. Let
$G(V,E)$
be a weighted graph designed to represent the information flow. Then, $G$ is in
fact a directed acyclic graph, with a set of vertices $V$ and a set of arcs
$E$.
The set $V$ is composed by three kinds of vertices:
\begin{itemize}
 \item Source vertex $S$: it represents the file to be stored.
There is only one source vertex in the graph.
 \item Data collector vertex $DC$: it represents the user who is allowed to
access the data in order to reconstruct the file.
 \item Storage node vertices $v_{in}^i$ and $v_{out}^i$: each storage node
$s_i$, where $i=1,\ldots, \infty$, is represented by one inner vertex $v_{in}^i$
and one outer vertex $v_{out}^i$.
\end{itemize}
In general, there is an arc $(v,w) \in E$ of weight $c$ from vertex
$v \in V$ to vertex $w \in V$ if $v$ can send $c$ data units to $w$.

At the beginning of the life of a distributed storage environment, there is
a file to be stored in $n$ storage nodes $s_i$, $i=1,\ldots,n$. This
can be represented by a source vertex $S$ with outdegree $n$ connected to
vertices $v_{in}^i$, $i=1,\ldots,n$. Since we are interested in analyzing the information
flow graph $G$ in terms of $\alpha$ and $\beta$, and these $n$ arcs are not
significant to find the mincut of $G$, their weight is set to infinite.
Moreover, to represent that each one of the storage nodes $s_i$, $i=1,\ldots,n$, stores $\alpha$ data units,
each vertex $v_{in}^i$ is connected to the vertex
$v_{out}^i$ with an arc of weight $\alpha$.

When the first storage node fails, the first newcomer  $s_{n+1}$ connects
to $d$ existing storage nodes sending, each one of them, $\beta$ data units.
This can be represented by adding one arc from $v_{out}^i$, $i=1,\ldots,n$, to $v_{in}^{n+1}$ of weight
$\beta$ if $s_i$ sends $\beta$ data units to $s_{n+1}$ in the regenerating
process. The new vertex $v_{in}^{n+1}$ is also connected to its associated
vertex $v_{out}^{n+1}$ with an arc of weight $\alpha$. This process can be
repeated for every failed node. Let the newcomers be denoted by $s_j$, where
$j=n+1,\ldots,\infty$.

Finally, after some failures, a data collector wants to reconstruct the file.
Therefore, a vertex $DC$ is added to $G$ along with one arc from
vertex $v_{out}^i$ to $DC$ if the data collector connects to the storage node
$s_i$. Note that if $s_i$ has been replaced by $s_j$, the
vertex $DC$ can not connect to $v_{out}^i$, but it can connect to
$v_{out}^j$. The vertex $DC$ has indegree $k$ and each arc has weight infinite,
because they have no relevance in finding the mincut of $G$.

If the mincut from vertex $S$ to $DC$, denoted by $\mbox{mincut}(S,DC)$,
achieves
that $\mbox{mincut}(S,DC)\ge M$, the data collector can reconstruct the file,
since there is enough information flow from the source to the data collector.
In fact, the data collector can connect to any $k$ nodes, so
$\min(\mbox{mincut}(S,DC))\ge
M$, which is achieved when the data collector connects to $k$ storage nodes that
have been already replaced by a newcomer \cite{Di01}. From this scenario, the
mincut is computed and lower bounds on the parameters $\alpha$ and $\gamma$ are
given. Let $\alpha^*(d,\gamma)$ be the threshold function, which is the function
that minimizes $\alpha$. Since $\alpha \ge \alpha^*(d,\gamma)$, if
$\alpha^*(d,\gamma)$ can be achieved, then any $\alpha \ge \alpha^*(d,\gamma)$
is also achieved.

Figure \ref{Fig:01} illustrates the information flow graph $G$ associated
to a $[4,2,3]$ regenerating code. Note that $\mbox{mincut}(S,DC) = \min
( 3\beta, \alpha ) + \min ( 2\beta, \alpha)$ which is the minimum mincut
for this information flow graph.
In general, it can be claimed that $\mbox{mincut}(S,DC)
\ge \sum_{i=0}^{k-1} \min ( (d-i)\beta,
\alpha ) \ge M$, which after an optimization process leads to the
following threshold function $\alpha^*(d,\gamma)$ also shown in \cite{Di01}:

\begin{equation}
\label{thresholdDim}
 \alpha^*(d,\gamma) = \left\{ \begin{array}{ll}
             \frac{M}{k}, & \gamma \in [f(0), +\infty) \\
             \\ \frac{M-g(i)\gamma}{k-i}, & \gamma \in [f(i),f(i-1)) \\
                & i=1, \ldots, k-1,
             \end{array}
   \right.
\end{equation}
where
$$f(i) = \frac{2Md}{(2k-i-1)i + 2k(d-k+1)} \text{ and }
g(i) = \frac{(2d-2k+i+1)i}{2d}.$$

Using the information flow graph $G$, we can see that there are exactly $k$
points in the tradeoff curve, or equivalently, $k$ intervals in the threshold
function $\alpha^*(d,\gamma)$, which represent $k$ newcomers. In the mincut
equation, the $k$ terms in the summation are computed as the minimum between two
parameters: the sum of the weights of the arcs that we have to cut to isolate
the corresponding $v^j_{in}$ from $S$, and the weight of the arc that we have to
cut to isolate the corresponding $v^j_{out}$ from $S$. Let the first parameter
be called the \textit{income} of the corresponding newcomer $s_j$. Note that the
income of the newcomer $s_j$ depends on the previous newcomers.

It can be seen that the newcomers can be ordered according to their income from
the highest to the lowest. In this model, this order is only determined by the
order of replacement of the failed nodes. Moreover, the MSR point corresponds to
the lowest income, which is given by the last newcomer added to the information
flow graph; and the MBR point corresponds to the highest, which is given by the
first newcomer. It is important to note also that, in this model, the order of
replacement of the nodes does not affect to the final result, since the mincut
is always the same independently of the specific set of $k$ failed nodes.

\subsection{Static cost model}
\label{subsec:02}

In \cite{Ak01}, \textit{Akhlaghi et al.} presented another distributed storage
model, where the storage nodes are partitioned into two sets $V^1$ and
$V^2$. Let $V^1$ be the set of ``cheap bandwidth'' nodes, from where each data
unit sent costs $C_c$, and $V^2 $ be the set of ``expensive bandwidth'' nodes,
from where each data unit sent costs $C_e$ such that $C_e > C_c$. This means
that when a newcomer replaces a lost storage node, the cost of downloading data
from a node in $V^1$ will be lower than the cost of downloading the same amount
of data from a node in $V^2$.

Consider the same situation as in the model described in Subsection
\ref{subsec:01}. Now, when a storage node fails, the newcomer node
$s_j$, $j=n+1,\ldots,\infty$, connects to $d_c$ existing storage nodes from
$V^1$ sending each one of them $\beta_c$ data units to $s_j$, and to $d_e$
existing storage nodes from $V^2$ sending each one of them $\beta_e$ data units
to $s_j$. Let $d=d_c+d_e$ be the number of helper nodes. Assume that $d$, $d_c$,
and $d_e$ are fixed, that is, they do not depend on the newcomer $s_j$,
$j=n+1,\ldots,\infty$. In terms of the information flow graph $G$, there is one
arc from $v_{out}^i$ to $v_{in}^j$ of weight $\beta_c$ or $\beta_e$, depending
on whether $s_i$ sends $\beta_c$ or $\beta_e$ data units, respectively, in the
regenerating process. This new vertex $v_{in}^j$ is also connected to its
associated vertex $v_{out}^j$ with an arc of weight $\alpha$.

Let the repair cost be $C_T = d_c C_c \beta_c + d_e C_e \beta_e$ and the repair
bandwidth $\gamma = d_c \beta_c + d_e \beta_e$. To simplify the model, we can
assume, without loss of generality, that $\beta_c = \tau \beta_e$ for some real
number $\tau \ge 1$. This means that we can minimize the repair cost $C_T$ by
downloading more data units from the set of ``cheap bandwidth'' nodes $V^1$ than
from the set of ``expensive bandwidth'' nodes $V^2$. Note that if $\tau$ is
increased, the repair cost is decreased and vice-versa.

Again, it must be satisfied that $\min(\mbox{mincut}(S,DC))\ge M$.
Moreover, the newcomers can also be ordered according to their income from the
highest to the lowest. However, in this model, the order is not only determined
by the order of replacement of the failed nodes, as it happened in the model
described in Subsection \ref{subsec:01}. It is important to note that, in this
model, the order of replacement of the nodes affects to the final result. The
mincut is not always the same, since it depends on the specific set of failed
nodes.

The goal is also to find the $\min(\mbox{mincut}(S,DC))$, so the next
problem arises: which is the set of $k$ newcomers that minimize the mincut
between $S$ and $DC$? The minimum mincut is given by the set of $k$ newcomers
with the minimum sum of incomes. As it is shown in \cite{Ak01}, this set is
composed by any $d_c+1$ newcomers from $V^1$ plus the remaining newcomers from
$V^2$. Moreover, the MSR point corresponds to the lowest income, which is
given by the last newcomer; and the MBR point corresponds to the highest income,
which is given by the first newcomer. Depending on $k$ and $d_c$, it is necessary to
distinguish between two cases.

\subsubsection{Case $k \le d_c+1$}

\begin{figure}
\centering
\begin{center}
\begin{tabular}{ll}
 \includegraphics[width=7.3cm, height=6cm]{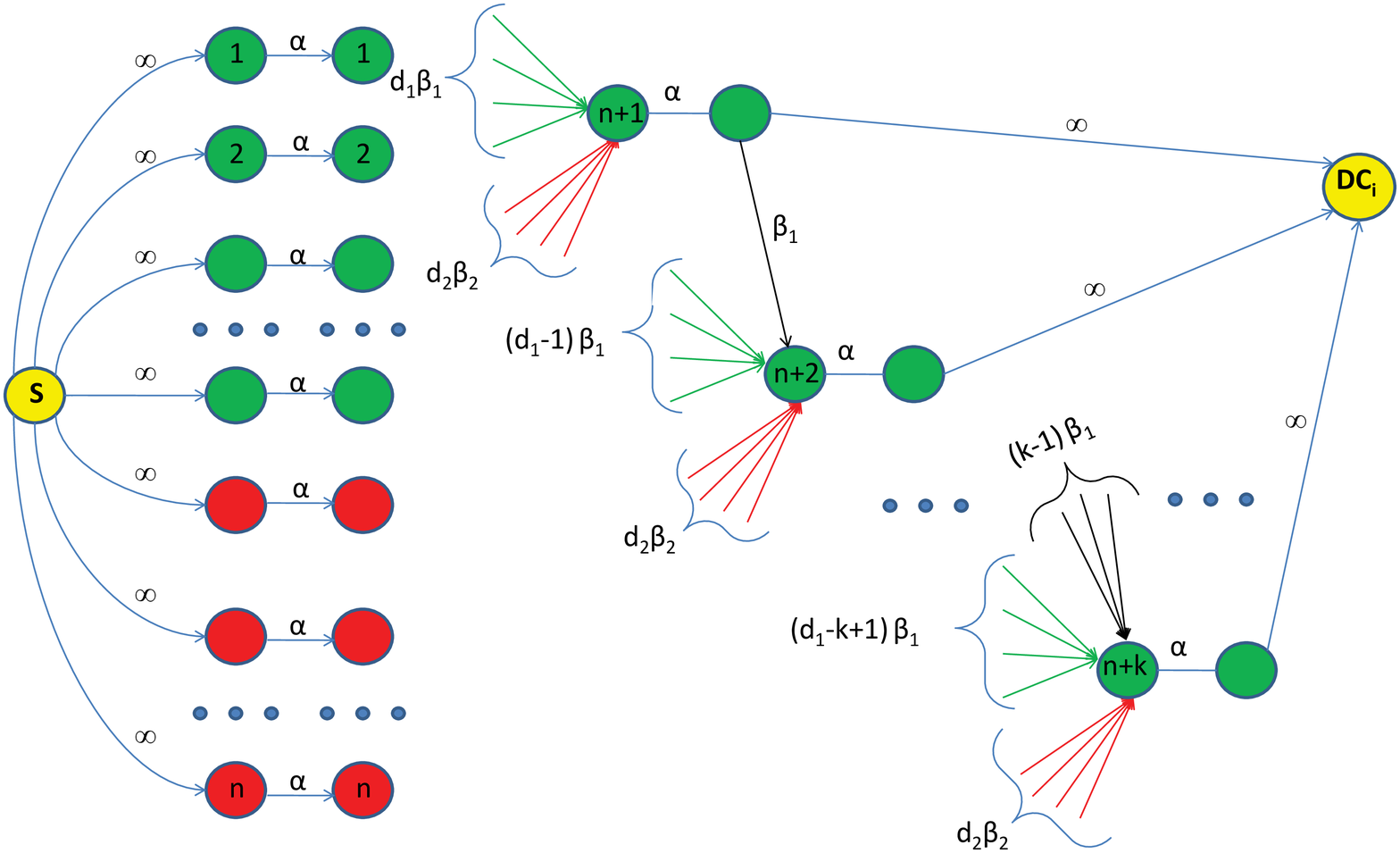}
&
 \includegraphics[width=7.3cm, height=6cm]{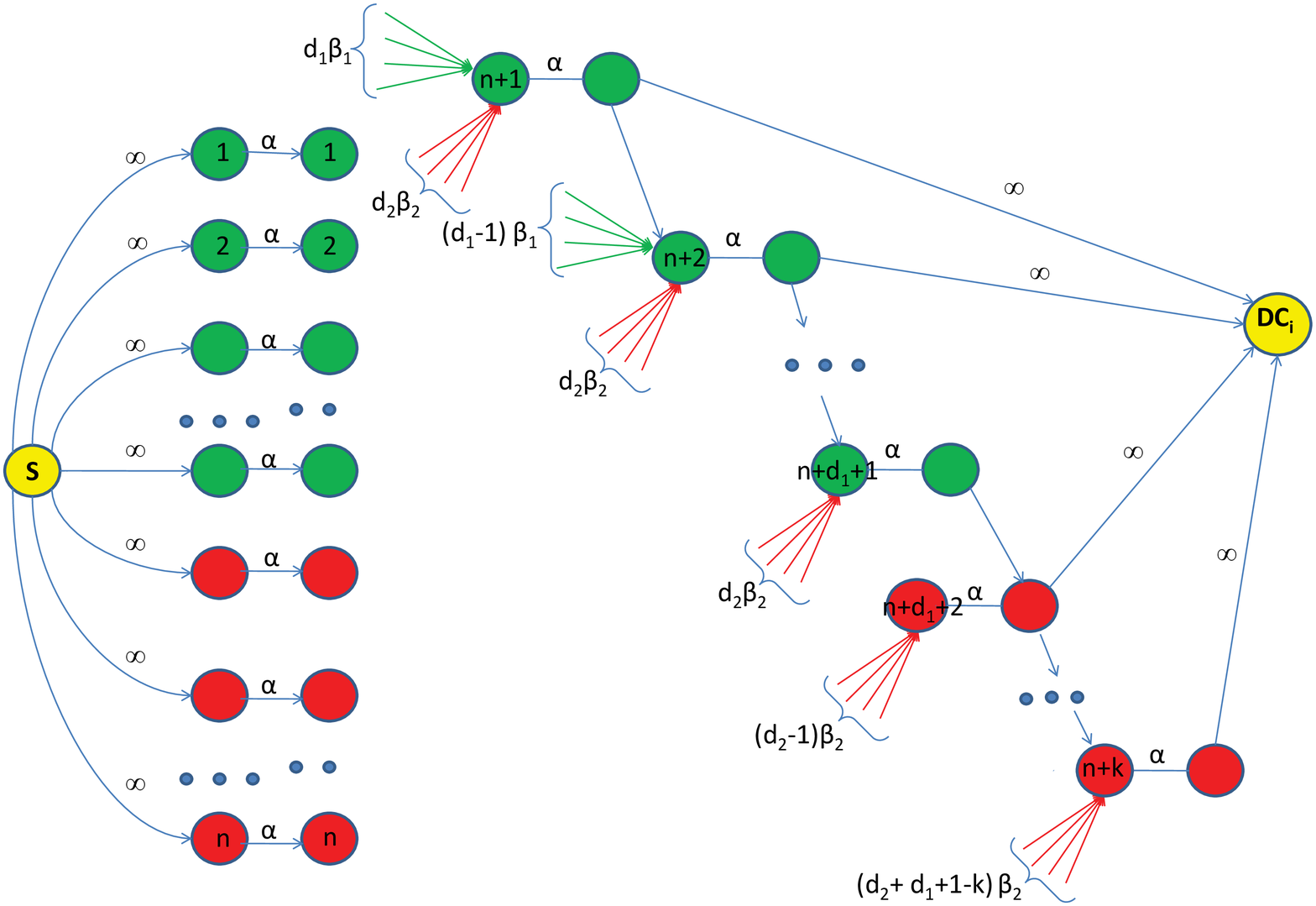}
\end{tabular}
\end{center}
\caption{ General information flow graphs corresponding to the cases $k
\le d_c+1$ (left) and $k > d_c+1$ (right).}
\label{Fig:02}
\end{figure}

This case corresponds to the situation when the data collector
connects to $k$ newcomers from the set $V^1$. With this scenario shown
in the information flow graph of Figure \ref{Fig:02} left, the mincut
analysis leads to
\begin{equation} \label{eq3}
 \sum_{i=0}^{k-1} \min ( {d_c\beta_c + d_e \beta_e -
i\beta_c , \alpha})  \ge M.
\end{equation}

After applying $\beta_c = \tau \beta_e$ and an optimization process,
the mincut equation (\ref{eq3}) leads to the following threshold function:

\begin{equation}
\label{ak_treshold_c}
 \alpha^*(d_c,d_e,\beta_e) = \left\{ \begin{array}{ll}
             \frac{M}{k}, & \beta_e \in [f(0), +\infty) \\
             \\ \frac{2M-g(i)\beta_e}{2(k-i)}, & \beta_e \in [f(i),f(i-1))
             \\ & i=1, \ldots, k-1,
             \end{array}
   \right.
\end{equation}
where
$$f(i)= \frac{2M}{2k(d_c \tau + d_e - \tau k ) + \tau (i + 1)(2k - i) } \;
\text{
and}$$
$$g(i) = i(2d_c \tau + 2d_e - 2k\tau + (i + 1)\tau).$$

\subsubsection{Case $k > d_c+1$}

This case corresponds to the situation when the data collector connects to
$d_c+1$
replaced nodes from the set $V^1$ and to $k-d_c-1$ replaced nodes from the set
$V^2$. With this scenario shown in the information flow graph of Figure
\ref{Fig:02} right, the mincut analysis leads to

\begin{equation}
\label{mincut:Ak}
\sum_{i=0}^{d_c} \min ( {d_c \beta_c + d_e \beta_e - i\beta_c ,
\alpha})  + \sum_{i=d_c+1}^{k-1} \min ( {(d_c + d_e -
i)\beta_e , \alpha}) \ge M.
\end{equation}

After applying $\beta_c = \tau \beta_e$ and an optimization process,
the mincut equation (\ref{mincut:Ak}) leads to the following threshold function:
\begin{equation}
 \alpha^*(d_c,d_e,\beta_e) = \left\{ \begin{array}{ll}
             \frac{M}{k}, & \beta_e \in [f_1(0), +\infty) \\
             \\ \frac{2M-g(i)\beta_e}{2(k-i)}, & \beta_e \in [f_1(i),f_1(i-1))
\\		& i = 1, \ldots, k-d_c-1 \\
	     \\ \frac{2M-(g_1(i)(k-d_c-1)g_2(i))\beta_e}{2(d_c-i)}, & \beta_e
\in [f_2(i),f_2(i-1)),
\\		& i = k-d_c, \ldots, k-1,
             \end{array}
   \right.
\end{equation}
where
$$f_1(i)= \frac{2M}{2k(d-k) + (i + 1) + (2k-1) },$$
$$f_2(i)= \frac{2M}{(2kd-k^2- d_c^2 - d_c + k +2d_c \tau) + i\tau(2d_c-i-1)},$$
$$g_1(i) = i(2d- 2k + i + 1), \text{ and}$$
$$g_2(i) = (i+1)(2d_e + i\tau).$$

\section{Two-rack model}
\label{sec:2}

In this model, the cost of sending data to a newcomer in a
different rack is higher than the cost of sending data to a newcomer in the same
rack. Note the difference of this rack model compared with the static cost model
described in Subsection \ref{subsec:02}. In that model, there is a static
classification of the storage nodes between the ones having ``cheap bandwidth''
and the ones having ``expensive bandwidth''. In our new model, this
classification depends on each newcomer. When a storage
node fails and a newcomer enters into the system, nodes from the same rack are
 in the ``cheap bandwidth'' set, while nodes in other racks are in the
``expensive bandwidth'' set. In this section, we analyze the case
when there are only two racks. Let $V_1$ and $V_2$ be the sets of $n_1$ and $n_2$ storage nodes
from the first and second rack, respectively.

Consider the same situation as in Subsection \ref{subsec:02}, but now the sets
of ``cheap bandwidth'' and ``expensive bandwidth'' nodes depend on the
specific replaced node. Again, we can assume, without loss of generality, that
$\beta_c = \tau \beta_e$ for some real number $\tau \ge 1$. Let the newcomers be
the storage nodes $s_j$, $j=n+1,\ldots, \infty$. Let $d= d_c^1+d_e^1= d_c^2
+d_e^2$ be the number of helper nodes for any newcomer, where $d_c^1$, $d_e^1$
and $d_c^2$, $d_e^2$ are the number of cheap and expensive bandwidth helper nodes
of a newcomer in the first and second rack, respectively. We can always
assume that $d_c^1 \le d_c^2$, by swapping racks if it is necessary.

In the model described in Subsection \ref{subsec:01}, the repair bandwidth
$\gamma$ is the same for
any newcomer. In the rack model, it depends on the rack where the newcomer is placed. Let
$\gamma^1=  \beta_e(d_c^1 \tau +d_e^1)$ be the repair bandwidth for any
newcomer in the first rack with repair cost $C_T^1 = \beta_e (C_c d_c^1 \tau +
C_e d_e^1)$, and let $\gamma^2= \beta_e (d_c^2 \tau +d_e^2)$ be the repair
bandwidth for any newcomer in the second rack with repair cost $C_T^2 = \beta_e
(C_c d_c^2 \tau + C_e d_e^2)$. Note that if $d_c^1=d_c^2$ or $\tau=1 $, then
$\gamma^1 = \gamma^2$, otherwise $\gamma^1 < \gamma^2$. As it is mentioned in
\cite{Di01}, in order to represent a distributed storage system, the information
flow graph is restricted to $\gamma \ge \alpha$. In the rack model,
it is necessary that $\gamma^1 \ge \alpha$, which means that
$\gamma^2 \ge \alpha$.

Moreover, unlike the models described in Section \ref{sec:1}, where it is
straightforward to establish which is the set of nodes which minimize the
mincut, in the rack model, this set of nodes may change depending on the
parameters $k$, $d_c^1$, $d_c^2$, $n_1$ and $\tau$. We call to this set
of newcomers, the \textit{minimum mincut set}. Recall that the income of a
newcomer $s_j$, $j=n+1,\ldots,\infty$, is the sum of the
weights of the arcs that should be cut in order to isolate $v_{in}^j$ from $S$.
Let $I$ be the indexed multiset containing the incomes of $k$ newcomers which
minimize the mincut. It is easy to see that in the model described in
Subsection \ref{subsec:01}, $I=\{ (d-i)\beta \;|\; i=0,\ldots, k-1 \}$,
and in the one described in Subsection \ref{subsec:02}, $I=\{ ((d_c-i) \tau
 +d_e) \beta_e \;|\; i=0,\ldots, \min (d_c, k-1) \} \cup \{ (d_e-i) \beta_e
\;|\; i=1,\ldots, k-d_c-1 \}$. Note that when $k\le d_c+1$, $\{ (d_e-i) \beta_e
\;|\; i=1,\ldots, k-d_c-1 \}$ is empty.

In order to establish $I$ in the rack model, the set of $k$ newcomers which
minimize the mincut must be found. First, note that since $d_c^1 \le d_c^2$, the
income of the newcomers is minimized by replacing first $d_c^1+1$ nodes from the
rack with less number of helper nodes, which in fact minimizes the
mincut. Therefore, the indexed multiset $I$ always contains the incomes of
a set of $d_c^1+1$ newcomers from $V_1$. Define $I_1=\{((d_c^1-i)\tau +
d_e^1)\beta_e \;| \; i=0,\ldots, \min (d_c^1, k-1) \} $ as the indexed multiset
where $I_1[i]$, $i=0,\ldots, \min (d_c^1, k-1 )$, are the incomes of this set of
$d_c^1+1$ newcomers from $V^1$. If $k \le d_c^1+1$, then $I=I_1$, otherwise $I_1
\subset I$ and $k-d_c^1-1$ more newcomers which minimize the mincut must be
found.

When $k > d_c^1+1$, we will see that there are two possibilities, either the
remaining nodes from $V_1$ are in the set of newcomers which minimize the mincut
or not. Define $I_2 = \{ d_e^1 \beta_e \;|\; i=1,\ldots, \min (k-d_c^1-1,
n_1-d_c^1-1 )\} \cup \{ (d_c^2-i) \tau \beta_e \;|\; i = 0,\ldots, \min (d_c^2,
k-n_1-1) \}$ as the indexed multiset where $I_2[i]$, $i=0,\ldots, k-d_c^1-2$, are
the incomes of a set of $k-d_c^1-1$ newcomers, including the remaining
$n_1-d_c^1-1$ newcomers from $V_1$ and newcomers from $V_2$. Note that if
$n_1-d_c^1-1 > k-d_c^1-1$, it only contains newcomers from $V_1$. Define $I_3 =
\{ (d_c^2-i) \tau \beta_e \; |\;i=0,\ldots, \min (d_c^2, k-d_c^1-2) \}$ as the
indexed multiset where $I_3[i]$, $i=0,\ldots, k-d_c^1-2$, are the incomes of a
set of $k-d_c^1-1$ newcomers from $V_2$. When $d_c^2 < k-d_c^1-1$ or
$d_c^2<k-n_1$, according to the information flow graph, the remaining incomes
necessary to complete the set of $k-d_c^1-1$ newcomers are zero. Therefore, it
can be assumed that $d_c^2 \ge k-d_c^1-1 \ge k-n_1$, since the mincut equation
does not change when $d_c^2 < k-d_c^1-1$ or $d_c^2<k-n_1$.

\begin{prop}
\label{prop:01}
If $k > d_c^1+1$, we have that $|I_2| = |I_3| = k-d_c^1-1$. Moreover, if
$\sum_{i=0}^{k-d_c^1-2} I_2[i] < \sum_{i=0}^{k-d_c^1-2} I_3[i]$, then $I=I_1
\cup I_2$; otherwise $I=I_1 \cup I_3$.
\end{prop}

\begin{demo}
We need to prove that $I_2$ and $I_3$ are the only possible sets of incomes
which minimize the mincut. We will see that it is not possible to find a set of
incomes such that the sum of all its elements is less than $\min (\sum_{i=0}^{|I_2|-1} I_2[i],
\sum_{i=0}^{|I_3|-1} I_3[i])$.

Let $A= I_2 - (I_2 \cap I_3) = \{ a_1, a_2, \ldots, a_n  \;|\; a_i=a_j,
i<j \}$ and $B= I_3 - (I_2 \cap I_3) = \{ b_1,b_2,\ldots, b_n \;|\; b_i>b_j, i<j
\}$. Let $D= A \cup B= \{d_1,d_2,\ldots, d_{2n} \;|\; d_i \ge d_j, i<j \}$. Then,
$\sum_{i=1}^n d_i \ge \sum_{i=1}^n  b_i$ and $\sum_{i=1}^n d_i\ge \sum_{i=1}^n
a_i$. Note that $A$, $B$ and $D$ are incomes of an information
flow graph, which means that one can not add $d_2$ without having added $d_1$
to the sum. The same happens with $A$ or $B$, so the elements must be included in
order from the highest to the lowest.
\end{demo}

\medskip
If $k\le d_c^1+1$, $I=I_1$ and the corresponding mincut equation is
\begin{equation}
\label{mincut1}
\sum_{i=0}^{|I_1|-1} \min
( I_1[i], \alpha ) \ge M.
\end{equation}
On the other hand, if $k>d_c^1+1$ and $I = I_1 \cup I_2$, the corresponding
mincut equation is
\begin{equation}
\label{mincut2}
\sum_{i=0}^{|I_1|-1} \min
( I_1[i], \alpha ) + \sum_{i=0}^{|I_2|-1} \min
( I_2[i], \alpha) \ge M,
\end{equation}
and if $I = I_1 \cup I_3$, the equation is
\begin{equation}
\label{mincut3}
\sum_{i=0}^{|I_1|-1} \min ( I_1[i], \alpha)
+ \sum_{i=0}^{|I_3|-1} \min ( I_3[i], \alpha) \ge M.
\end{equation}

In the previous models described in Section \ref{sec:1}, the decreasing behavior
of the incomes included in the mincut equation is used to find the threshold
function which minimizes the parameters $\alpha$ and $\gamma$.  In the rack
model, the incomes included in the mincut equations may not have a decreasing
behavior as the newcomers enter into the system, so it is necessary to find the
threshold function in a different way.

Let $L$ be the increasing ordered list of values such that for all $i, \; i=
0,\ldots, k-1$, $I[i]/\beta_e \in L$ and $|I|=|L|$.
Note that any of the information flow graphs, which represent the
rack model or any of the two models from Section \ref{sec:1}, can be described
in terms of $I$, so they can be represented by $L$. Therefore, once $L$ is
found, it is possible to find the parameters $\alpha$ and $\beta_e$ (and then
$\gamma$ or $\gamma^1$ and $\gamma^2$) using the threshold function
given in the next theorem. Note that the way to represent this threshold
function for the rack model can be seen as a generalization, since it also
represents the behavior of the mincut equations for the previous given models.

\begin{theo}
\label{theo:1}
 The threshold function $\alpha^*(\beta_e)$ (which also depends on
$d$, $d_c^1$, $d_c^2$, $k$ and $\tau$) is the following:
\begin{equation}
\label{treshold3}
 \alpha^*(\beta_e) = \left\{ \begin{array}{ll}
             \frac{M}{k}, & \beta_e \in [f(0), +\infty)  \\ & \\
              \frac{M-g(i)\beta_e}{k-i}, & \beta_e \in
[f(i),f(i-1))\\ &
i=1, \ldots, k-1,
   \end{array}
   \right.
\end{equation}
subject to $\gamma^1=(d_c^1 \tau + d_e^1) \beta_e \ge \alpha$,
where
$$f(i) = \frac{M}{L[i](k-i)+g(i)} \; \text{ and } \;
g(i) =\sum_{j=0}^{i-1} L[j].$$
Note that $f(i)$ is a decreasing function and $g(i)$ is an increasing function.
\end{theo}

\begin{demo}
We want to obtain the threshold function which minimizes $\alpha$, that is,
\begin{equation}
\label{opti3}
   \begin{array}{ll} \alpha^*(\beta_e) = & \min \alpha \\ &
\text{subject to: } \sum_{i=0}^{k-1} \min ( L[i]\beta_e, \alpha) \ge M.
\end{array}
\end{equation}
Therefore, we are going to show the optimization of (\ref{opti3}) which
leads to the threshold function (\ref{treshold3}).

Define $M^*$ as $$ M^*  = \sum_{i=0}^{k-1} \min ( L[i] \beta_e, \alpha). $$
Note that $M^*$ is a piecewise linear function of $\alpha$. Since $L$ is a
sorted list of $k$ values, if $\alpha$ is less than the lowest value $L[0]$,
then $M^*=k\alpha$. As $\alpha$ grows, the values from $L$ are added to the
equation, so
\begin{equation}
\label{func:01}
 M^* = \left\{ \begin{array}{ll}
             k \alpha, & \alpha \in [0, L[0]\beta_e] \\
	 \\ (k-i) \alpha + \sum_{j=0}^{i-1} L[j] \beta_e, & \alpha \in
(L[i-1]\beta_e, L[i]\beta_e] \\ & i=1,\ldots,k-1\\
	 \\ \sum_{j=0}^{k-1} L[j] \beta_e, & \alpha \in
(L[k-1]\beta_e, \infty).
   \end{array}
   \right.
\end{equation}

Using that $M^* \ge M$, we can minimize $\alpha$ depending on $M$. Note that
the term $\sum_{j=0}^{k-1} L[j] \beta_e$ of the previous equation has no
significance in the minimization of $\alpha$, so it can be ignored. Therefore,
we obtain the function

\begin{equation}
\label{func:02}
 \alpha^* = \left\{ \begin{array}{ll}
             \frac{M}{k}, & M \in [0, k L[0] \beta_e] \\
	 \\ \frac{M - \sum_{j=0}^{i-1} L[j] \beta_e}{k-i}, & M \in
(L[i-1]\beta_e (k-i) + \sum_{j=0}^{i-1} L[j] \beta_e, \\
	 & L[i]\beta_e (k-i) + \sum_{j=0}^{i-1} L[j] \beta_e] \\
	& i=1,\ldots,k-1.
   \end{array}
   \right.
\end{equation}

Finally, define $g(i) =\sum_{j=0}^{i-1} L[j]$ and $f(i) = \frac{M}{L[i](k-i)+g(i)}$.
Then, the above expression of $\alpha^*$ can be defined over $\beta_e$ instead of
over $M$, and the threshold function (\ref{treshold3}) follows.
\end{demo}

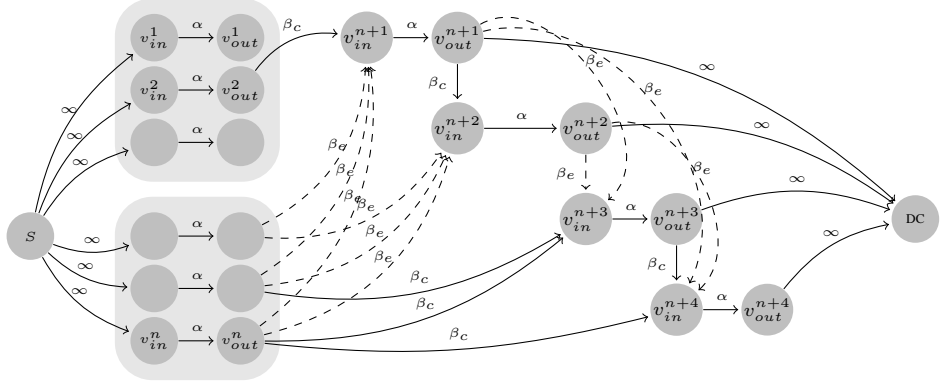
\begin{figure}[ht]
\centering
\begin{tikzpicture}[shorten >=1pt,->,font= \tiny]
  \tikzstyle{vertix}=[circle,fill=black!25,minimum size=18pt,inner
sep=0pt,font=\tiny]
  \tikzstyle{invi}=[circle]
  \tikzstyle{background}=[rectangle, fill=gray!20, inner sep=0.2cm,rounded
corners=5mm]

  \node[vertix] (s) {$S$};

  \node[vertix, right= 1cm of s] (vin_4)  {};

  \node[vertix, above=0.6cm of vin_4] (vin_3)  {};
  \node[vertix, below=0.05 of vin_4] (vin_5)  {};

  \node[vertix, above=0.05 of vin_3] (vin_2)  {$v_{in}^2$};
  \node[vertix, below=0.05 of vin_5] (vin_6)  {$v_{in}^n$};

  \node[vertix,above=0.05 of vin_2] (vin_1)  {$v_{in}^1$};

 \foreach \to in {1,2,3}
   {\path (s) edge[bend left=20, font= \tiny] node[anchor=south,above]{$\infty$}
(vin_\to);}
 \foreach \to in {4,5,6}
   {\path (s) edge[bend right=20, font= \tiny]
node[anchor=south,above]{$\infty$}
 (vin_\to);}

  \node[vertix, right= 0.5cm of vin_1] (vout_1)  {$v_{out}^1$};
  \node[vertix, right= 0.5cm of vin_2] (vout_2)  {$v_{out}^2$};
  \node[vertix, right= 0.5cm of vin_3] (vout_3)  {};
  \node[vertix, right= 0.5cm of vin_4] (vout_4)  {};
  \node[vertix, right= 0.5cm of vin_5] (vout_5)  {};
  \node[vertix, right= 0.5cm of vin_6] (vout_6)  {$v_{out}^n$};

  \node[vertix, right= 1cm of vout_1] (vin_7)
{\scriptsize{$v_{in}^{n+1}$}};
  \node[vertix, right= 0.5 cm of vin_7] (vout_7)
{\scriptsize{$v_{out}^{n+1}$}};

  \node[vertix, below= 0.5cm of vout_7]
(vin_8)
{\scriptsize{$v_{in}^{n+2}$}};
  \node[vertix, right= 1cm of vin_8] (vout_8)
{\scriptsize{$v_{out}^{n+2}$}};

  \node[vertix, below= 0.5cm of vout_8]
(vin_9)
{\scriptsize{$v_{in}^{n+3}$}};
  \node[vertix, right=0.5cm of vin_9] (vout_9)
{\scriptsize{$v_{out}^{n+3}$}};

  \node[vertix, below=0.5cm of vout_9]
(vin_10)
{\scriptsize{$v_{in}^{n+4}$}};
  \node[vertix, right=0.5cm of vin_10] (vout_10)
{\scriptsize{$v_{out}^{n+4}$}};

  \node[vertix, right=2.5cm of vout_9] (DC)
{DC};

 \foreach \to in {7,8,9,10}
   {\path (vout_\to) edge[bend left=20, font= \tiny]
node[anchor=south,above]{$\infty$}
 (DC);}

  \path[->, bend left, font= \tiny] (vout_2) edge
node[anchor=south,above]{$\beta_c$} (vin_7);
  \path[dashed,->, bend right] (vout_4) edge
node[anchor=south,above]{$\beta_e$}(vin_7);
  \path[->, dashed, bend right] (vout_5) edge
node[anchor=south,above]{$\beta_e$} (vin_7);
  \path[->, dashed, bend right] (vout_6) edge
node[anchor=south,above]{$\beta_e$} (vin_7);

  \path[->] (vout_7) edge node[anchor=south,left]{$\beta_c$}
(vin_8);
  \path[->, dashed, bend right] (vout_4) edge
node[anchor=south,above]{$\beta_e$} (vin_8);
  \path[->, dashed, bend right] (vout_5) edge
node[anchor=south,above]{$\beta_e$} (vin_8);
  \path[->, dashed, bend right] (vout_6) edge
node[anchor=south,above]{$\beta_e$} (vin_8);

  \path[->, dashed, bend left=85] (vout_7) edge node[anchor=south,
above]{$\beta_e$}
(vin_9);
  \path[->,dashed] (vout_8) edge node[anchor=south,left]{$\beta_e$}
(vin_9);
  \path[->, bend right=20] (vout_5) edge
node[anchor=south,above]{$\beta_c$} (vin_9);
  \path[->, bend right=20] (vout_6) edge
node[anchor=south,above]{$\beta_c$} (vin_9);

  \path[->, bend left=65, dashed, font= \tiny] (vout_7) edge node[anchor=south,
above]{$\beta_e$}
(vin_10);
  \path[->, bend left=75, dashed, font= \tiny] (vout_8) edge
node[anchor=south,above]{$\beta_e$}
(vin_10);
  \path[->] (vout_9) edge node[anchor=south,left]{$\beta_c$}
(vin_10);
  \path[->, bend right=10, font= \tiny] (vout_6) edge
node[anchor=south,above]{$\beta_c$} (vin_10);

 \foreach \from/\to in {1,2,3,4,5,6,7,8,9,10}
  { \path[->, font= \tiny] (vin_\from) edge node[anchor=south] {$\alpha$}
(vout_\to); }

\begin{pgfonlayer}{background}
 \node [background,fit= (vin_1) (vin_3) (vout_3)] {};
 \node [background,fit= (vin_4) (vin_6) (vout_6)] {};
\end{pgfonlayer}
\end{tikzpicture}
\caption{Information flow graph corresponding to the rack model
when $k > d_c^1+1$,
with $k=4$, $d_c^1=1$, $d_c^2=2$, $d=4$ and $n_1=n_2=3$.
}
\label{Fig:03}
\end{figure}

\begin{exmp}
Figure \ref{Fig:03} shows the example of an information flow graph
corresponding
to a regenerating code with $k=4$, $d_c^1=1$, $d_c^2=2$, $d=4$ and $n_1=n_2=3$.
Taking for example $\tau= 2$, we have that $I_1=\{5\beta_e,3\beta_e \}$, $I_2=
\{3\beta_e,4 \beta_e \}$ and $I_3=\{4\beta_e,2\beta_e \}$. By Proposition
\ref{prop:01}, since $\sum_{i=0}^{1} I_2[i] > \sum_{i = 0}^{1}I_3[i]$, $I= I_1
\cup I_3 = \{5\beta_e,3\beta_e,4\beta_e,2\beta_e\}$, and then $L=[2,3,4,5]$.
The corresponding mincut equation is (\ref{mincut3}) and applying
$L$ to the threshold function (\ref{treshold3}), we obtain
\begin{equation}
 \alpha^*(\beta_e) = \left\{ \begin{array}{ll}
             \frac{M}{4}, & \beta_e \in [\frac{M}{8}, +\infty)  \\ &\\
             \frac{M-2 \beta_e}{3}, & \beta_e \in
[\frac{M}{11},\frac{M}{8})\\ & \\
	       \frac{M-5 \beta_e}{2}, & \beta_e \in
[\frac{M}{13},\frac{M}{11})\\ & \\
	       M-9 \beta_e, & \beta_e \in [\frac{M}{14},\frac{M}{13}).\\
   \end{array}
   \right.
\end{equation}

\end{exmp}

It can happen that two consecutive values in $L$ are equal, that is
$L[i]=L[i-1]$, so $f(i) = f(i-1)$. In this case, we consider that the interval
$[f(i),f(i-1))$ is empty and it can be deleted.

\begin{exmp}
Figure \ref{Fig:04} shows the same example as Figure \ref{Fig:03} with an
information flow graph corresponding to a regenerating code with $d_c^1=1$,
$d_c^2=2$, $d=4$ and $n_1=n_2=3$, but taking $k=3$ instead of $k=4$. If
for example $\tau= 2$, we have that $I_1=\{ 5\beta_e,3\beta_e \}$, $I_2=\{3\beta_e\}$ and
$I_3=\{4\beta_e\}$. By Proposition \ref{prop:01}, since $\sum_{i=0}^{0}
I_2[i] < \sum_{i=0}^{0} I_3[i]$, $I= I_1 \cup I_2 = \{ 5\beta_e,3\beta_e
, 3\beta_e\}$, and then $L=[3,3,5]$. The corresponding mincut equation is
(\ref{mincut2}) and applying $L$ to the threshold function (\ref{treshold3}), we
obtain

\begin{equation}
 \alpha^*(\beta_e) = \left\{ \begin{array}{ll}
             \frac{M}{3}, & \beta_e \in [\frac{M}{9}, +\infty)  \\
             \\ \frac{M-3 \beta_e}{2}, & \beta_e \in
[\frac{M}{9},\frac{M}{9})\\
	      \\ M-6 \beta_e, & \beta_e \in [\frac{M}{11},\frac{M}{9}).
   \end{array}
   \right.
\end{equation}
Note that the second interval is empty and it can be deleted.

\end{exmp}
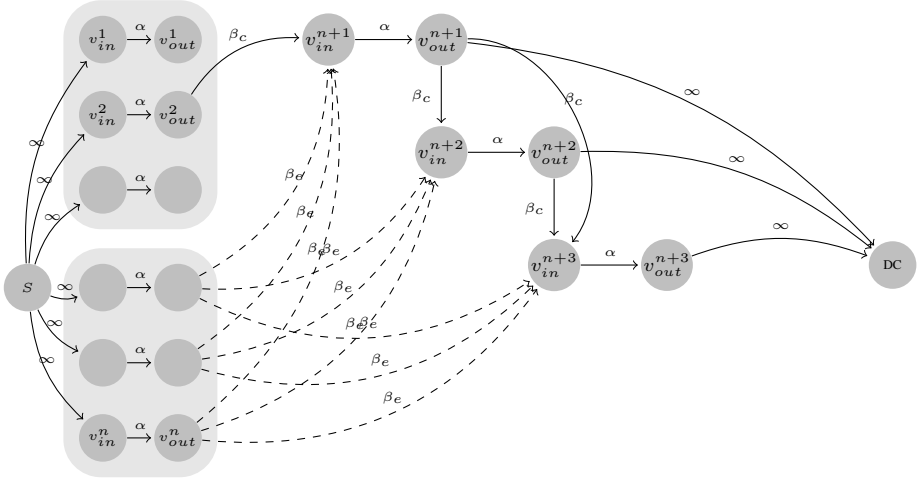
\begin{figure}
\centering
\begin{tikzpicture}[shorten >=1pt,->,font= \tiny]
  \tikzstyle{vertix}=[circle,fill=black!25,minimum size=18pt,inner
sep=0pt,font=\tiny]
  \tikzstyle{invi}=[circle]
  \tikzstyle{background}=[rectangle, fill=gray!20, inner sep=0.2cm,rounded
corners=5mm]

  \node[vertix] (s) {$S$};

  \node[vertix, right of=s] (vin_4)  {};

  \node[vertix, above of=vin_4, node distance=1.3cm] (vin_3)  {};
  \node[vertix, below of=vin_4] (vin_5)  {};

  \node[vertix, above of=vin_3] (vin_2)  {$v_{in}^2$};
  \node[vertix, below of=vin_5] (vin_6)  {$v_{in}^n$};

  \node[vertix,above of=vin_2] (vin_1)  {$v_{in}^1$};

 \foreach \to in {1,2,3}
   {\path (s) edge[bend left=20] node[anchor=south,above]{$\infty$}
(vin_\to);}
 \foreach \to in {4,5,6}
   {\path (s) edge[bend right=20]
node[anchor=south,above]{$\infty$}
 (vin_\to);}

  \node[vertix, right of=vin_1] (vout_1)  {$v_{out}^1$};
  \node[vertix, right of=vin_2] (vout_2)  {$v_{out}^2$};
  \node[vertix, right of=vin_3] (vout_3)  {};
  \node[vertix, right of=vin_4] (vout_4)  {};
  \node[vertix, right of=vin_5] (vout_5)  {};
  \node[vertix, right of=vin_6] (vout_6)  {$v_{out}^n$};

  \node[vertix, right of=vout_1, node distance = 2cm] (vin_7)
{\scriptsize{$v_{in}^{n+1}$}};
  \node[vertix, right of=vin_7, node distance = 1.5cm] (vout_7)
{\scriptsize{$v_{out}^{n+1}$}};

  \node[vertix, below of= vout_7, node distance = 1.5cm]
(vin_8)
{\scriptsize{$v_{in}^{n+2}$}};
  \node[vertix, right of=vin_8, node distance = 1.5cm] (vout_8)
{\scriptsize{$v_{out}^{n+2}$}};

  \node[vertix, below of= vout_8, node distance = 1.5cm]
(vin_9)
{\scriptsize{$v_{in}^{n+3}$}};
  \node[vertix, right of=vin_9, node distance = 1.5cm] (vout_9)
{\scriptsize{$v_{out}^{n+3}$}};

  \node[vertix, right of=vout_9, node distance = 3cm] (DC)
{DC};

 \foreach \to in {7,8,9}
   {\path (vout_\to) edge[bend left=20]
node[anchor=south,above]{$\infty$}
 (DC);}

  \path[->, bend left] (vout_2) edge
node[anchor=south,above]{$\beta_c$} (vin_7);
  \path[dashed,->, bend right] (vout_4) edge
node[anchor=south,above]{$\beta_e$}(vin_7);
  \path[->, dashed, bend right] (vout_5) edge
node[anchor=south,above]{$\beta_e$} (vin_7);
  \path[->, dashed, bend right] (vout_6) edge
node[anchor=south,above]{$\beta_e$} (vin_7);

  \path[->] (vout_7) edge node[anchor=south,left]{$\beta_c$}
(vin_8);
  \path[->, dashed, bend right] (vout_4) edge
node[anchor=south,above]{$\beta_e$} (vin_8);
  \path[->, dashed, bend right] (vout_5) edge
node[anchor=south,above]{$\beta_e$} (vin_8);
  \path[->, dashed, bend right] (vout_6) edge
node[anchor=south,above]{$\beta_e$} (vin_8);

  \path[->, bend left=65] (vout_7) edge node[anchor=south,
above]{$\beta_c$}
(vin_9);
  \path[->] (vout_8) edge node[anchor=south,left]{$\beta_c$}
(vin_9);
 \path[->, dashed, bend right] (vout_4) edge
node[anchor=south,above]{$\beta_e$} (vin_9);
  \path[->, dashed, bend right] (vout_5) edge
node[anchor=south,above]{$\beta_e$} (vin_9);
  \path[->, dashed, bend right] (vout_6) edge
node[anchor=south,above]{$\beta_e$} (vin_9);
;

 \foreach \from/\to in {1,2,3,4,5,6,7,8,9}
  { \path[->] (vin_\from) edge node[anchor=south] {$\alpha$}
(vout_\to); }

\begin{pgfonlayer}{background}
 \node [background,fit= (vin_1) (vin_3) (vout_3)] {};
 \node [background,fit= (vin_4) (vin_6) (vout_6)] {};
\end{pgfonlayer}
\end{tikzpicture}
\caption{ Information flow graph corresponding to the rack model
when $k > d_c^1+1$,
with $k=3$, $d_c^1=1$, $d_c^2=2$, $d=4$ and $n_1=n_2=3$.
}
\label{Fig:04}
\end{figure}

\medskip
Finally, note that when $k \le d_c^1+1$, the mincut equations and the threshold function
(\ref{treshold3}) for the rack model are exactly the same as the ones
shown in \cite{Ak01} for the model described in Subsection \ref{subsec:02}.
Actually, it can be seen that $d_c^1$ of the rack model is equivalent to $d_c$ of
the static cost model. Indeed, it can be seen that when $k \le d_c^1+1$, the rack
model and the static cost model have the same behavior because $I=I_1$.

\subsection{MSR and MBR points}
\label{subsec:03}

The threshold function (\ref{treshold3}) leads to a tradeoff curve
between $\alpha$ and $\beta_e$. Note that, like in the static cost model,
since there is a different repair bandwidth $\gamma^1$ and $\gamma^2$ for each
rack,
this curve is based on $\beta_e$ instead of $\gamma^1$ and $\gamma^2$.

At the MSR point, the amount of stored data per node is $\alpha_{MSR} = M/k$.
Moreover, at this point, the minimum value of $\beta_e$ is $\beta_e = f(0) =
\frac{M}{L[0]k}$, which leads to $$\gamma^1_{MSR} = \frac{(d_c^1 \tau +
d_e^1)M}{L[0]k} \quad \textrm{and} \quad  \gamma^2_{MSR} = \frac{(d_c^2 \tau +
d_e^2)M}{L[0]k}.$$
On the other hand, at the MBR point, as $f(i)$ is a
decreasing function, the parameter $\beta_e$ which leads to the minimum
repair bandwidths is $\beta_e = f(|L|-1) = \frac{M}{L[|L|-1](k-|L|+1)+g(|L|-1)}
$. Then,
the corresponding amount of stored data per node is $\alpha_{MBR} =
\frac{M L[|L|-1]}{(k-|L|+1) L[|L|-1] + g(|L|-1)}$, and the repair
bandwidths
are $$\gamma^1_{MBR} = \frac{(d_c^1 \tau + d_e^1)M}{L[|L|-1](k-|L|+1)+g(|L|-1)}
\quad \textrm{and}$$
$$\gamma^2_{MBR} = \frac{(d_c^2 \tau + d_e^2)M}{L[|L|-1](k-|L|+1)+g(|L|-1)}.$$

\subsection{Non-feasible situation}
\label{subsec:04}

As we have seen, the threshold function (\ref{treshold3}) is subject to $\gamma^1=(d_c^1 \tau +
d_e^1) \beta_e \ge \alpha$.

\begin{prop}
\label{prop:02}
If the inequality $\gamma^1 \ge \alpha$ is achieved, then $\max (L) =
I_1[0]/\beta_e$.
\end{prop}
\begin{demo}
Since $L$ is an increasing ordered list, for $i=0,\ldots,k-1$, $\max
(L) = L[k-1]$. As $I_1[0]$ is the income of the first newcomer, then $I_1[0]/
\beta_e = d_c^1 \tau + d_e^1 \in L$. Actually, $L$ is constructed from all elements
in $I$ and $I_1\subseteq I$, by Proposition \ref{prop:01}.

If $\gamma^1\geq \alpha$, then taking $\beta_e=f(k-1)$ in Theorem
\ref{theo:1}, we have that
$\gamma^1= (d_c^1 \tau + d_e^1) \beta_e= (d_c^1 \tau + d_e^1) f(k-1)\geq
M-g(k-1)f(k-1)$. After some operations, we obtain that
$\frac{ (d_c^1 \tau+d_e^1)M }{ \sum_{j=0}^{k-1} L[j] } \ge
\frac{L[k-1]M}{\sum_{j=0}^{k-1} L[j]}$, so $d_c^1 \tau+d_e^1 \ge L[k-1]$.
Since $I_1[0]/\beta_e = d_c^1 \tau + d_e^1 \in L$ and $\max (L) = L[k-1]$,
$d_c^1 \tau+d_e^1 = L[k-1] = I_1[0]/\beta_e$.
\end{demo}

\medskip
Since any distributed storage system satisfies that $\gamma^1 \ge \alpha$,
we have that $\max (L) = I_1[0]/\beta_e$, by Proposition \ref{prop:02}.
In order to have this situation, we need to
remove from $L$ any value $L[i]$ such that $L[i] >
I_1[0]/\beta_e$, $i=0,\ldots,k-1$. After that, we can assume that $L[|L|-1] =
I_1[0]/\beta_e$. In terms of the tradeoff curve, this means that there is no
point in the curve that outperforms the MBR point.

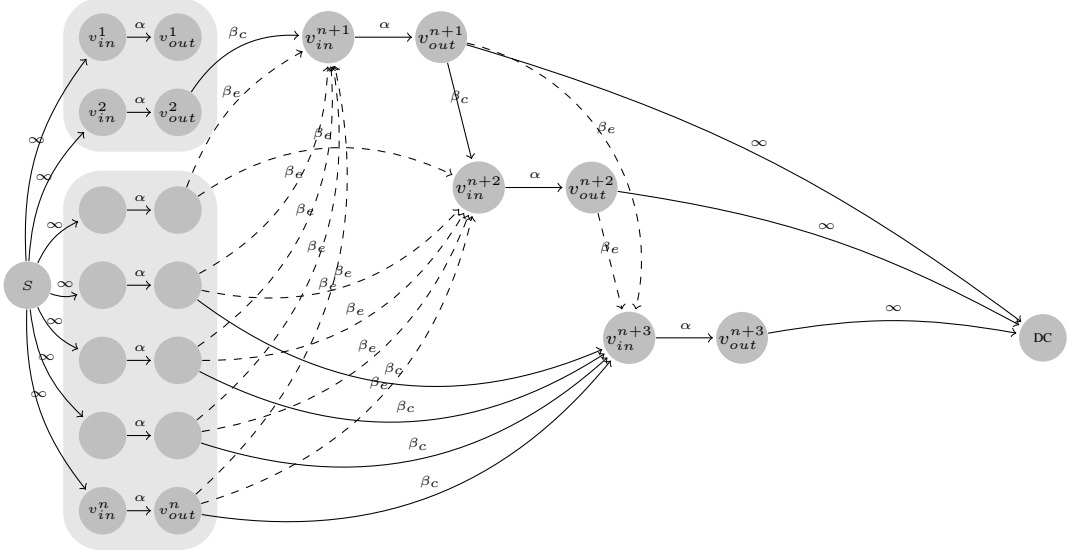
\begin{figure}
\centering
\begin{tikzpicture}[shorten >=1pt,->,font= \tiny]
  \tikzstyle{vertix}=[circle,fill=black!25,minimum size=18pt,inner
sep=0pt,font=\tiny]
  \tikzstyle{invi}=[circle]
  \tikzstyle{background}=[rectangle, fill=gray!20, inner sep=0.2cm,rounded
corners=5mm]

  \node[vertix] (s) {$S$};

  \node[vertix, right of=s] (vin_4)  {};
  \node[vertix, above of=vin_4] (vin_3)  {};
  \node[vertix, below of=vin_4] (vin_5)  {};
  \node[vertix, above of=vin_3, node distance=1.3cm] (vin_2)  {$v_{in}^2$};
  \node[vertix, below of=vin_5] (vin_6)  {};
  \node[vertix, above of=vin_2] (vin_1)  {$v_{in}^1$};
  \node[vertix, below of=vin_6] (vin_l)  {$v_{in}^n$};

 \foreach \to in {1,2,3}
   {\path (s) edge[bend left=20] node[anchor=south,above]{$\infty$} (vin_\to);}
 \foreach \to in {4,5,6,l}
   {\path (s) edge[bend right=20]  node[anchor=south,above]{$\infty$}
 (vin_\to);}

  \node[vertix, right of=vin_1] (vout_1)  {$v_{out}^1$};
  \node[vertix, right of=vin_2] (vout_2)  {$v_{out}^2$};
  \node[vertix, right of=vin_3] (vout_3)  {};
  \node[vertix, right of=vin_4] (vout_4)  {};
  \node[vertix, right of=vin_5] (vout_5)  {};
  \node[vertix, right of=vin_6] (vout_6)  {};
  \node[vertix, right of=vin_l] (vout_l)  {$v_{out}^n$};

  \node[vertix, right of=vout_1, node distance = 2cm] (vin_7)
{\scriptsize{$v_{in}^{n+1}$}};
  \node[vertix, right of=vin_7, node distance = 1.5cm] (vout_7)
{\scriptsize{$v_{out}^{n+1}$}};

  \node[vertix, right of=vout_2, below of= vin_7, node distance = 2cm]
(vin_8)
{\scriptsize{$v_{in}^{n+2}$}};
  \node[vertix, right of=vin_8, node distance = 1.5cm] (vout_8)
{\scriptsize{$v_{out}^{n+2}$}};

  \node[vertix, right of=vout_3, below of= vin_8, node distance = 2cm]
(vin_9)
{\scriptsize{$v_{in}^{n+3}$}};
  \node[vertix, right of=vin_9, node distance = 1.5cm] (vout_9)
{\scriptsize{$v_{out}^{n+3}$}};

  \node[vertix, right of=vout_9, node distance = 4cm] (DC)
{DC};

 \foreach \to in {7,8,9}
   {\path (vout_\to) edge[bend left=10]  node[anchor=south,above]{$\infty$}
 (DC);}

  \path[->, bend left] (vout_2) edge
node[anchor=south,above]{$\beta_c$} (vin_7);
  \path[->, dashed, bend left=20] (vout_3) edge
node[anchor=south,above]{$\beta_e$}(vin_7);
  \path[->, dashed, bend right] (vout_4) edge
node[anchor=south,above]{$\beta_e$}(vin_7);
  \path[->, dashed, bend right] (vout_5) edge
node[anchor=south,above]{$\beta_e$} (vin_7);
  \path[->, dashed, bend right] (vout_6) edge
node[anchor=south,above]{$\beta_e$} (vin_7);
  \path[->, dashed, bend right] (vout_l) edge
node[anchor=south,above]{$\beta_e$} (vin_7);
  \path[->] (vout_7) edge node[anchor=south,above]{$\beta_c$} (vin_8);
  \path[->, dashed, bend left] (vout_3) edge
node[anchor=south,above]{$\beta_e$} (vin_8);
  \path[->, dashed, bend right] (vout_4) edge
node[anchor=south,above]{$\beta_e$} (vin_8);
  \path[->, dashed, bend right] (vout_5) edge
node[anchor=south,above]{$\beta_e$} (vin_8);
  \path[->, dashed, bend right] (vout_6) edge
node[anchor=south,above]{$\beta_e$} (vin_8);
  \path[->, dashed, bend right] (vout_l) edge
node[anchor=south,above]{$\beta_e$} (vin_8);

  \path[->, bend left=45, dashed] (vout_7) edge
node[anchor=south,above]{$\beta_e$}
(vin_9);
  \path[->,dashed] (vout_8) edge node[anchor=south,above]{$\beta_e$}
(vin_9);
 \path[->, bend right] (vout_4) edge
node[anchor=south,above]{$\beta_c$} (vin_9);
  \path[->, bend right] (vout_5) edge
node[anchor=south,above]{$\beta_c$} (vin_9);
  \path[->, bend right] (vout_6) edge
node[anchor=south,above]{$\beta_c$} (vin_9);
  \path[->, bend right] (vout_l) edge
node[anchor=south,above]{$\beta_c$} (vin_9);

 \foreach \from/\to in {1,2,3,4,5,6,7,8,9,l}
  { \path[->] (vin_\from) edge node[anchor=south] {$\alpha$} (vout_\to); }

\begin{pgfonlayer}{background}
 \node [background,fit= (vin_1) (vin_2) (vout_2)] {};
 \node [background,fit= (vin_3) (vin_l) (vout_l)] {};
\end{pgfonlayer}
\end{tikzpicture}
\caption{Information flow graph with $k=3$, $n_1=2$, $n_2=5$,
$d_c^1=1$, $d_c^2=4$ and $d=6$.
}
\label{Fig:05}
\end{figure}

\begin{exmp}
In order to illustrate this situation, we can consider the example
of a regenerating code with $k=3$, $d_c^1=1$, $d_c^2=4$, $d=6$, $n_1=2$ and $n_2=5$,
and the information flow graph given in Figure \ref{Fig:05}.
Taking $\tau=2$, the incomes of the newcomers $s_{n+1}$, $s_{n+2}$ and $s_{n+3}$
are $7 \beta_e$, $5 \beta_e$ and $8 \beta_e$, respectively.
Actually, we have that $I=I_1 \cup I_2$, where $I_1= \{7\beta_e,5\beta_e \}$ and $I_2=\{8\beta_e \}$.
Then, $L=[5,7,8]$, so $\max{(L)}=8 > I[0]/\beta_e=7$. Applying $L$ to the
threshold function (\ref{treshold3}), the resulting minimization of $\alpha$ and
$\beta_e$ is
$$ \alpha^*(\beta_e) = \left\{ \begin{array}{ll}
             \frac{M}{3}, & \beta_e \in [\frac{M}{15}, +\infty)\\
             \\ \frac{M-5\beta_e}{2}, & \beta_e \in
[\frac{M}{19},\frac{M}{15})\\
	     \\M-12 \beta_e, & \beta_e \in [\frac{M}{20},\frac{M}{19}).
   \end{array}
   \right.
$$

Note that considering the last interval, we have that for $\beta_e=f(k-1)=\frac{M}{20}$,
$\alpha_{MBR}=\frac{8M}{20}$  and $\gamma^1_{MBR} = (d_c^1 \tau +
d_e^1) f(k-1) = \frac{7M}{20}$. Applied to the information flow graph,
we obtain that $\mbox{mincut}(S,DC) = \frac{7M}{20}+ \frac{5M}{20}+
\frac{8M}{20} = M$ which is true. However, since $\alpha_{MBR} > \gamma^1_{MBR}$,
it gives a non-feasible situation for a distributed storage scheme. Note also that if we
delete this non-feasible interval, then $\gamma^1_{MBR}= \frac{7M}{19}$ and
$\alpha_{MBR} = \frac{7M}{19}$ which corresponds to the MBR point because $\gamma^1_{MBR} =
\alpha_{MBR}$.
\end{exmp}

It is important to note that more than one element from $I$ can be greater
than any element from $I_1$, which will result in more impossible intervals. In
conclusion, any value from $I$ greater than the greatest value from $I_1$,
must be deleted because otherwise it would lead to a non-feasible situation.

\subsection{Case $d_e^1 \beta_e \ge d_c^2 \tau \beta_e$}

In this case, the mincut equation has a decreasing behavior as $i$ increases
for $i=0,\ldots,k-1$. Therefore, it is possible to define an injective
function with a decreasing behavior, which will be used to determine the
intervals of the threshold function. Basically, it is possible to use the same
procedure shown in \cite{Di01} and \cite{Ak01} to find the threshold function.
Moreover, it can be seen that the set of incomes which minimize the mincut is
always the same, it does not depend on any parameter.

It is easy to see that if $d_e^1 \beta_e \ge d_c^2 \tau \beta_e$ and $k \le d_c^1+1$, the mincut equations (and so
the threshold functions) corresponding to the model explained in this section and the model
explained in Subsection \ref{subsec:02} are exactly the same. Therefore, we
will focus on the situation that $d_e^1 \beta_e \ge d_c^2 \tau \beta_e$ and $k > d_c^1+1$.
Note that this is in fact a particular case of the
general threshold function (\ref{treshold3}), where it
is possible to create a decreasing function for any feasible $i$, and then
find the threshold function giving more details.

\begin{theo}
When $d_e^1 \ge d_c^2 \tau$ and $k > d_c^1+1$, the threshold function $\alpha^*(\beta_e)$ (which
also depends on $d$, $d_c^1$, $d_c^2$, $k$ and $\tau$) is the following:
\begin{equation}
\label{treshold2}
 \alpha^*(\beta_e) = \left\{ \begin{array}{ll}
             \frac{M}{k}, & \beta_e \in [f_1(0), +\infty)  \\
             \\ \frac{M-g_1(i)\tau\beta_e}{k-i}, & \beta_e \in
[f_1(i),f_1(i-1))\\ & i=1, \ldots, k-d_c^1-2 \\
	    \\ \frac{M-g_1(k-d_c^1-1)\tau\beta_e}{k-i}, & \beta_e \in
[f_2(k-d_c^1-1),f_1(k-d_c^1-2))\\
	    \\ \frac{M-g_1(k-d_c^1-1)\tau \beta_e - g_2(i-k+d_c^1+1)
\beta_e}{k-i}, & \beta_e \in [f_2(i), f_2(i-1))\\&
i=k-d_c^1, \ldots, k-1,
   \end{array}
   \right.
\end{equation}
where
$$g_1(i) =\frac{i}{2}(2d-2k+i+1),$$
$$g_2(i)= \frac{i}{2}(2d_e^1+\tau i-\tau), $$
$$f_1(i) = \frac{2M}{\tau(2k(d-k)+(i+1)(2k-i))}, \text{ and }$$
$$f_2(i) = \frac{2M}{2 d_e^1 + 2 d_e^1 d_c^1 - \tau (i
(i-2k+1)+2(k^2-k-kd+d_e^1+d_e^1 d_c^1))}.$$
Note that $f_1(i)$ and $f_2(i)$, $i=0,\ldots,k-1$, are decreasing functions,
and
$g_1(i)$ and $g_2(i)$, $i=1,\ldots,k-1$, are increasing functions.
\end{theo}

\begin{demo}
Note that $d_e^1 = d_c^2+1$ and $d_e^2 = d_c^1+1$. We consider the mincut
equation (\ref{mincut3}) of the rack model, since if $d_e^1  \ge d_c^2 \tau$,
then we have that $I=I_1\cup I_3$, by Proposition \ref{prop:01}. In other words,
the $n_1-d_c^1-1$ remaining newcomers from $V^1$ are not in the set of newcomers
which minimizes the mincut. Assume that
$k \le d = d_c^1+d_e^1$ because if $d < k$, requiring any $d$ storage nodes to have
a flow of $M$ will lead to the same condition as requiring any $k$ storage nodes
to have a flow of $M$ \cite{Di01}. We want to obtain the threshold function which minimizes
$\alpha$, that is,
\begin{equation}
\label{opti2}
   \begin{array}{ll} \alpha^*(\beta_e) = & \min \alpha \\ &
\text{subject to: } \sum_{i=0}^{d_c^1} \min ( d_c^1 \beta_c + d_e^1
\beta_e - i \beta_c, \alpha ) + \\ & \sum_{i= d_c^1+1}^{k-1} \min
((d_c^1+d_e^1-i) \beta_c, \alpha) \ge M.
\end{array}
\end{equation}
Therefore, we are going to show the optimization of (\ref{opti2}) which leads
to (\ref{treshold2}).

Applying that $\beta_c = \tau \beta_e$, we can define the minimum $M$ as $M^*$,
so
$$
 M^*  = \sum_{i=0}^{d_c^1} \min ( (d_c^1 \tau + d_e^1  - i
\tau) \beta_e, \alpha ) + \sum_{i= d_c^1+1}^{k-1} \min (
(d_c^1+d_e^1-i) \tau \beta_e, \alpha).
$$
In order to change the order of the above summation, we define $$b(i_1,i_2) =
d_c^1 + d_e^1 - k +1 +i_1 +i_2 \tau.$$ Note that $M^*$ is a piecewise linear
function of $\alpha$. The minimum value of $\lbrace(d_c^1 \tau + d_e^1  - i
\tau) \beta_e \;|\; i =0, \ldots, d_c^1\} \cup \{ (d_c^1+d_e^1-i) \tau \beta_e
\;|\; i=d_c^1+1,\ldots,k-1\rbrace$ is when $i=k-1$. Therefore, if $\alpha$ is
less than this value, then $M^* = k\alpha$. Since $d_e^1=d_c^2+1$ and $d_e^2=
d_c^1+1$ the lowest value of $\{(d_c^1 \tau + d_e^1  - i \tau) \beta_e \;|\;
i =0,\ldots,d_c^1\}$ which is $d_e^1 \beta_e$, is higher than or equal to the
highest value of $\{(d_c^1+d_e^1-i) \tau \beta_e \;|\; i= d_c^1+1, \ldots,
k-1\}$, which is $(d_e^1-1) \tau \beta_e$. This means that as $\alpha$
increases, the term $(d_c^1+d_e^1-i) \tau \beta_e$ is added more times in $M^*$
while $i=k-1, \ldots, d_c^1$. When $i=d_c^1,\ldots, 0$, the term $(d_c^1 \tau +
d_e^1  - i \tau) \beta_e$ is added more times in $M^*$.

\begin{equation}
\label{eq:01}
 M^* = \left\{ \begin{array}{ll}
             k \alpha, & \alpha \in [0, b(0,0)\tau \beta_e] \\ & \\
	  \\ (k-i) \alpha + \sum_{j=0}^{i-1} b(j,0) \tau \beta_e, & \alpha
\in (b(i-1,0)\tau \beta_e, b(i,0) \tau \beta_e] \\
 & i=1, \ldots, k-d_c^1-2 \\
	 \\ (d_c^1+1)\alpha +  \sum_{j=0}^{k-d_c^1-2} b(j,0) \tau \beta_e, &  \alpha
\in (b(k-d_c^1-2,0)\tau \beta_e,\\ & b(k-d_c^1-1,0) \beta_e]\\&\\

	 \\ (k-i) \alpha + \sum_{j=0}^{k-d_c^1-2} b(j,0) \tau \beta_e +\\
\sum_{j=0}^{i-k+d_c^1} b(k-d_c^1-1,j) \beta_e , & \alpha \in (b(k-d_c^1-1,i-k+d_c^1)
\beta_e,\\& b(k-d_c^1-1,i-k+d_c^1+1) \beta_e] \\
& i=k-d_c^1, \ldots, k-1
\\ \sum_{j=0}^{k-d_c^1-2} b(j,0) \tau \beta_e +\\

\sum_{j=0}^{d_c^1} b(k-d_c^1-1,j) \beta_e , & \alpha \in (b(k-d_c^1-1,d_c^1)
\beta_e, \infty). \\&
   \end{array}
   \right.
\end{equation}

Using that $M \ge M^*$, we can minimize $\alpha$ depending on $M$. Note that the
last term of (\ref{eq:01}) does not affect in the minimization of
$\alpha$, so it is ignored. Therefore, we obtain the function
\begin{equation}
\label{eq:02}
 \alpha^* = \left\{ \begin{array}{ll}
             \frac{M}{k}, & M \in [0, k b(0,0) \tau \beta_e] \\
	 \\ \frac{M - \sum_{j=0}^{i-1} b(j,0) \tau \beta_e}{k-i}, & M \in
(A(i-1),A(i)] \\ & i=1, \ldots, k-d_c^1-2  \\
	 \frac{M - \sum_{j=0}^{i-1} b(j,0) \tau \beta_e}{k-i}, & M \in
(A(i-1),B(i)] \\
      \\ \frac{M-\sum_{j=0}^{k-d_c^1-2} b(j,0) \tau \beta_e -
\sum_{j=0}^{i-k+d_c^1} b(k-d_c^1-1,j) \beta_e}{k-i}, & M \in ( B(i-1), B(i)] \\&
i=k-d_c^1, \ldots, k-1,
   \end{array}
   \right.
\end{equation}
where $A(i) = \tau \beta_e (b(i,0)(k-i-1) + \sum_{j=0}^{i} b(j,0))$ and
$ B(i) = \beta_e (b(k-d_c^1-1,i-k+d_c^1+1) (k-i-1) + \sum_{j=0}^{k-d_c^1-2} b(j,0)
\tau + \sum_{j=0}^{i-k+d_c^1+1} b(k-d_c^1-1,j))$.

From the definition of $b(i_1,i_2)$,
$$\sum_{j=0}^{i-1} b(j,0) = \frac{i}{2}(2d-2k+i+1) = g_1(i), $$
$$\sum_{j=0}^{i-1}b(k-d_c^1-1,j) =\frac{i}{2}(2d_e^1+\tau i-\tau)= g_2(i),$$
$$ \tau ((k-i-1) b(i,0) + \sum_{j=0}^{i}b(j,0)) =
\frac{2M}{\tau(2k(d-k)+(i+1)(2k-i))}=
\frac{M}{f_1(i)} $$
and
$$ b(k-d_c^1-1,i-k+d_c^1+1) (k-i-1) + \sum_{j=0}^{k-d_c^1-2} b(j,0) \tau
+  \sum_{j=0}^{i-k+d_c^1+1} b(k-d_c^1-1,j)= \frac{M}{f_2(i)}.$$

The function (\ref{eq:02}) for $\alpha^*$ can be defined over
$\beta_e$ instead of over $M$, and then function (\ref{treshold2}) follows.
\end{demo}

\section{General rack model}
\label{sec:3}

Let $r \ge 2$ be the number of racks of a distributed storage system. Let $n_j$,
$j=1,\ldots,r$, be the number of storage nodes in the $j$-th rack. Let $d_c^j$ be the
number of helper nodes providing cheap bandwidth and $d_e^j$ be the number of
helper nodes providing expensive bandwidth to any newcomer in the $j$-th rack.
We assume that the total number of helper nodes $d$ is fixed, so it is satisfied
that $d= d_c^j+d_e^j$ for $j=1,\ldots,r$. Moreover, it can be seen that $d_e^j =
\sum_{z=1,z \ne  j}^r (d_c^z+1)$. Let the racks be increasingly ordered by number
of cheap bandwidth nodes, so $i \leq j$ if and only if $d_c^i \le d_c^j$.
First, we consider the case when $d=n-1$,
and then the general case, that is, when $d \le n-1$

\subsection{When $d=n-1$}
\label{subsec:05}

In this case, we impose that any available node
in the system is a helper node, that is, $d=n-1$.
If one node fails in the $j$-th rack, $d_c^j= n_j-1$ nodes from the
same rack and $d_e^j= n - n_j$ nodes from other racks help in the
regeneration process.

The indexed multiset $I$ containing the incomes of the $k$ newcomers
which minimize the mincut is
\begin{equation}
I=\bigcup_{j=1}^{r} \{ ((d_c^j - i) \tau + d_e^j - \sum_{z=1}^{j-1} (d_c^z - j
+1) \beta_e \; | \;
i=0,\ldots, \min(d_c^j,k-\sum_{z=1}^{j-1} d_c^z -j) \},
\label{eq:2}
\end{equation}
where $\sum_{z=1}^{0} x =0$ for any value $x$. Therefore, the resulting mincut
equation is $\sum_{i=0}^{k-1} \min(I[i],\alpha)\geq M$.

Finally, the threshold function (\ref{treshold3}) can be applied, so $\alpha$
and $\beta_e$ can be minimized.
Note that the set of $k$ newcomers which minimize the mincut is fixed
independently of $\tau$, so there is only one candidate set to be the minimum
mincut set.

\subsection{When $d \le n -1$}
\label{subsec:06}

In this case, there may exist nodes in the system that, after a node
failure, do not help in the regeneration process. These kind of systems
introduce
the difficulty of finding the minimum mincut set in the information flow graph.
Note that in the two-rack model,
after including the first $d_c^1+1$
nodes from the first rack, we need to known whether the remaining $n_1-d_c^1-1$
are included in the minimum mincut set or not. In order to solve this point,
we create two candidate sets to be the minimum mincut set, one with
these nodes and another one without them.

Define the indexed multiset $I'= \bigcup_{j=1}^r \{ (( d_c^j-i) \tau + d_e^j
- \sum_{z=1}^{j-1} d_c^z -j +1) \beta_e\; |\; i=0,\ldots,d_c^j \} \cup
I^j$, where $I^j= \{ (d_e^j- \sum_{z=1}^ {j-1} d_c^z -j +1)\beta_e \;| \;
i=1,\ldots, n_j-d_c^j-1 \}$
contains the incomes of the remaining $n_j-d_c^j-1$
newcomers once the first $d_c^j+1$ storage nodes have already been
replaced. Note that $I'$ represents the  incomes of all the $n$ newcomers.
Also note that in the $r$-th rack, $(d_e^r- \sum_{z=1}^{r-1} d_c^z -r +1)\beta_e
=0$,
and that Subsection \ref{subsec:05} describes the particular case when
$n_j- d_c^j-1 = 0$ for all $j=1,\ldots,r$.

We say that a rack is involved in the minimum mincut if at least one
of its nodes is in a candidate set to be the minimum mincut set.
The involved racks are always the first $s$ racks, where $s$
is the minimum number such that $\sum_{j=1}^s (d_c^j+1) \ge k$.
Since the newcomers corresponding to the incomes from $I^s$ are never included
in the minimum mincut set, the number of
candidate sets to be the minimum mincut set is $2^{s-1}$.  However, as the goal
is to find the set having the minimum sum of its corresponding incomes,
it is possible to design a linear algorithm with complexity $O(s-1)$ to
solve this problem. This algorithm is described in the next paragraph.


%
For all $j=1,\ldots,s-1$, if $\sum_{i=0}^{k-1} I'[i] > \sum_{i=0}^{k-1}
(I'-I^j)[i]$, where $I'-I^j$ means removing the elements of $I^j$ inside $I'$,
the new $I'$ becomes $I'-I^j$. This process is repeated for every $j$. Finally,
after $s-1$ comparisons, we obtain that $I=I'$. Then, we can assure that $I$
contains the incomes of the minimum mincut set of newcomers.
Once $I$ is found, we can define $L$ as in the two-rack model and apply the
threshold function (\ref{treshold3}) in order to minimize $\alpha$ and
$\beta_e$.

\begin{figure}
\centering
\begin{tikzpicture}[shorten >=1pt,->]
  \tikzstyle{vertix}=[circle,fill=black!25,minimum size=18pt,inner sep=0pt,
node distance = 0.8cm,font=\small]
  \tikzstyle{invi}=[circle]
  \tikzstyle{background}=[rectangle, fill=gray!20, inner sep=0.1cm,rounded
corners=5mm]

  \node[vertix] (v_1)  {};
  \node[vertix, below of=v_1] (v_2)  {};
  \node[vertix, below of=v_2] (v_3)  {};
  \node[vertix, below= 0.5cm of v_3] (v_4)  {};
  \node[vertix, below of=v_4] (v_5)  {};
  \node[vertix, below of=v_5] (v_6)  {};
  \node[vertix, below of=v_6] (v_7)  {};
  \node[vertix, below= 0.5cm of v_7] (v_8)  {};
  \node[vertix, below of=v_8] (v_9)  {};
  \node[vertix, below of=v_9] (v_10)  {};
  \node[vertix, below of=v_10] (v_11)  {};

  \node[vertix, right= 0.9cm of v_1] (vin_1)  {};
  \node[vertix, right= 0.7cm of vin_1] (vout_1)  {};
  \node[vertix, right of=v_2, below of=vout_1] (vin_2)  {};
  \node[vertix, right= 0.7cm of vin_2] (vout_2)  {};
  \node[vertix, right of=v_4, below= 1.2cm of vin_2] (vin_3)  {};
  \node[vertix, right= 0.7cm of vin_3] (vout_3)  {};
  \node[vertix, right of=v_5, below of=vout_3] (vin_4)  {};
  \node[vertix, right= 0.7cm of vin_4] (vout_4)  {};
  \node[vertix, right of=v_6, below of=vout_4] (vin_5)  {};
  \node[vertix, right= 0.7cm of vin_5] (vout_5)  {};
  \node[vertix, right of=v_7, below of=vout_5] (vin_6)  {};
  \node[vertix, right= 0.7cm of vin_6] (vout_6)  {};
  \node[vertix, right of=v_8, below= 0.6cm of vout_6] (vin_7)  {};
  \node[vertix, right= 0.7cm of vin_7] (vout_7)  {};

  \foreach \to in {1,2,3,4,5,6,7}
   {\path (vin_\to) edge  node[anchor=south,above]{$\alpha$}
 (vout_\to);}

    \foreach \to in {2}
   {\path (v_\to) edge  node[anchor=south,above] {$\tau \beta_e$}
 (vin_1);}
    \foreach \to in {4,5,6,8,9,10,11}
   {\path (v_\to) edge  node[anchor=south,above] {}
 (vin_1);}

    \foreach \to in {4,5,6,8,9,10,11}
   {\path (v_\to) edge  node[anchor=south,above] {}
 (vin_2);}
    \foreach \to in {1}
   {\path (vout_\to) edge  node[anchor=south,above] {$\tau \beta_e$}
 (vin_2);}

    \foreach \to in {5,6}
   {\path (v_\to) edge  node[anchor=south,above] {$\tau \beta_e$}
 (vin_3);}
    \foreach \to in {8,9,10,11}
   {\path (v_\to) edge  node[anchor=south,above] {}
 (vin_3);}
   \path[bend left=120] (vout_1) edge  node[anchor=south,above] {$\beta_e$}
 (vin_3);
    \path (vout_2) edge  node[anchor=south,above] {$\beta_e$}
 (vin_3);


   \path (v_6) edge  node[anchor=south,above] {$\tau \beta_e$}
 (vin_4);
    \foreach \to in {8,9,10,11}
   {\path (v_\to) edge  node[anchor=south,above] {}
 (vin_4);}
    \path (vout_3) edge  node[anchor=south,above] {$\tau\beta_e$}
 (vin_4);
    \foreach \to in {1,2}
   { \path[bend left=120] (vout_\to) edge  node[anchor=south,above] {$\beta_e$}
 (vin_4);}

    \foreach \to in {8,9,10,11}
   {\path (v_\to) edge  node[anchor=south,above] {}
 (vin_5);}
     \path (vout_4) edge  node[anchor=south,above] {$\tau\beta_e$}
 (vin_5);
      \path[bend left=80] (vout_3) edge  node[anchor=south,above]
{$\tau\beta_e$}
 (vin_5);
     \foreach \to in {1,2}
   { \path[bend left=120] (vout_\to) edge  node[anchor=south,above] {$\beta_e$}
 (vin_5);}

    \foreach \to in {8,9,10,11}
   {\path (v_\to) edge  node[anchor=south,above] {}
 (vin_6);}
      \path (vout_5) edge  node[anchor=south,above] {$\tau\beta_e$}
 (vin_6);
      \path[bend left=80] (vout_4) edge  node[anchor=south,above]
{$\tau\beta_e$}
 (vin_6);
      \path[bend left=80] (vout_3) edge  node[anchor=south,above]
{$\tau\beta_e$}
 (vin_6);
     \foreach \to in {1,2}
   { \path[bend left=120] (vout_\to) edge  node[anchor=south,above] {$\beta_e$}
 (vin_6);}

    \foreach \to in {9,10,11}
   {\path (v_\to) edge  node[anchor=south,above] {$\tau \beta_e$}
 (vin_7);}

    \foreach \to in {1,2,3,4,5}
   { \path[bend left=120] (vout_\to) edge  node[anchor=south,above] {$\beta_e$}
 (vin_7);}

\begin{pgfonlayer}{background}
 \node [background,fit= (v_1) (v_3) (vout_2)] {};
 \node [background,fit= (v_4) (v_7) (vout_6)] {};
 \node [background,fit= (v_8) (v_11) (vout_7)] {};
\end{pgfonlayer}

\end{tikzpicture}
\caption{ Information flow graph corresponding to the rack model with $k=7$,
$d_c^1=1$, $d_c^2=2$, $d_c^3=3$ and $d=8$.
}
\label{Fig:06}
\end{figure}
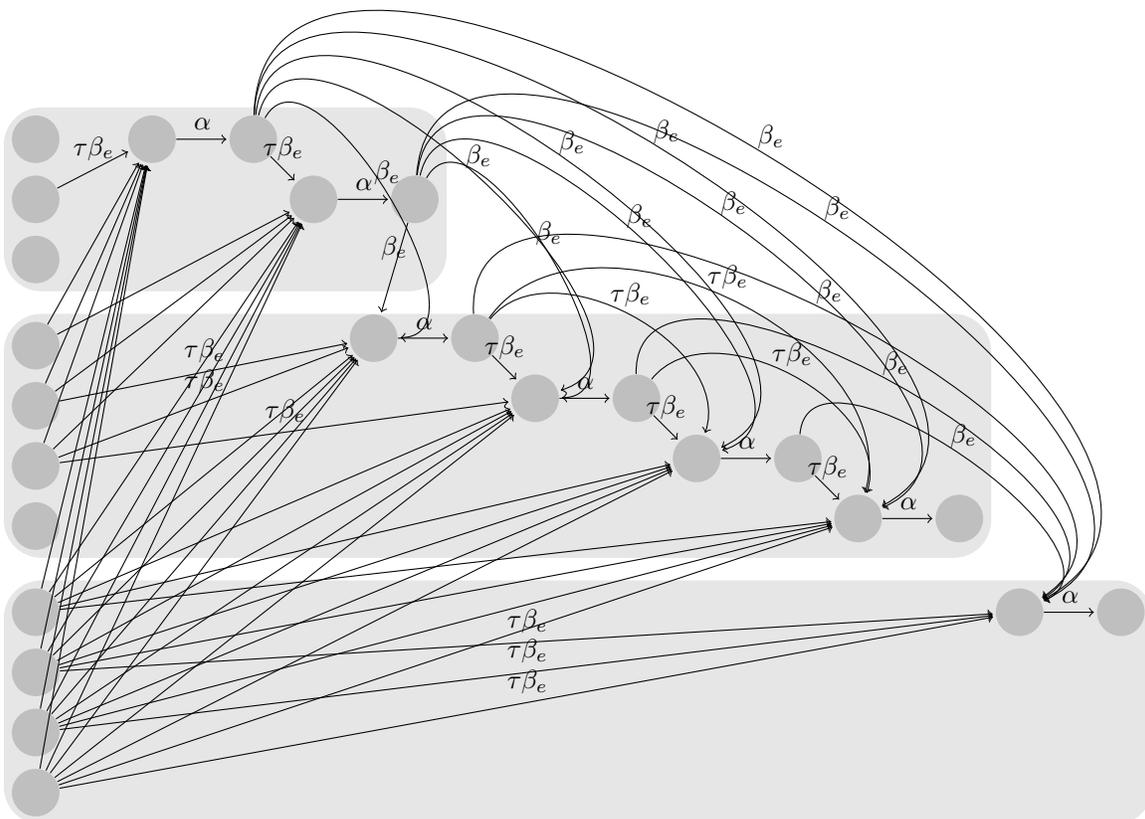

\begin{exmp}
Let the number of
racks be $r=3$ with $n_1=3$, $n_2=4$, $n_3=4$ and $k=7$. Let the number of
helper
nodes for any newcomer be $d=8$ with $d_c^1=1$, $d_c^2=2$ and $d_c^3=3$, so with
$d_e^1=7$, $d_e^2=6$ and $d_e^3=5$. Note that $d_c^1 \le d_c^2 \le d_c^3$. The
information flow graph corresponding to these parameters is shown in Figure
\ref{Fig:06}.

Since $s=3$, the three racks are involved in the minimum mincut and the incomes
in $I$ depend on whether the sets $I^1$ and $I^2$ are included or not:
\begin{itemize}
 \item Including $I^1$ and $I^2$: $I_{\{1,2\}}'=\{(\tau+7)\beta_e,7 \beta_e,
7\beta_e,
(2\tau+4)\beta_e, (\tau+4) \beta_e, 4 \beta_e, 4\beta_e\}$.
 \item Including $I^1$ but not $I^2$:
$I_{\{1\}}'=\{(\tau+7)\beta_e,7 \beta_e, 7\beta_e, (2\tau+4)\beta_e,
(\tau+4) \beta_e, 4 \beta_e, 3 \tau \beta_e\}$.
 \item Including $I^2$ but not $I^1$:
$I_{\{2\}}'=\{(\tau+7)\beta_e,7 \beta_e, (2\tau+4)\beta_e, (\tau+4) \beta_e,
4 \beta_e, 4\beta_e\,  3 \tau\beta_e \}$.
 \item Excluding $I^1$ and $I^2$:
$I_{\emptyset}'=\{(\tau+7)\beta_e,7 \beta_e, (2\tau+4)\beta_e, (\tau+4) \beta_e,
4
\beta_e, 3 \tau\beta_e , 2 \tau \beta_e \}$.
\end{itemize}
Then, if for example $\tau=2.2$, the sum of the elements of the above multisets
are $45.8\beta_e$, $48.4\beta_e$, $45.4\beta_e$ and $45.8\beta_e$, respectively.
So $I= I'_{\{2\}}$ contains the incomes corresponding to the minimum mincut set.

We can obtain the same result by using the algorithm proposed in this section,
that is, following these steps:
\begin{enumerate}
 \item Create $I'= \{(\tau+7)\beta_e,7 \beta_e, 7\beta_e, (2\tau+4)\beta_e,
(\tau+4) \beta_e, 4 \beta_e, 4\beta_e\, 3 \tau \beta_e, 2\tau \beta_e,
\tau\beta_e, 0 \}$.
 \item Create $I^1 = \{7 \beta_e \}$. Since $\sum_{i=0}^{6} I'[i] =
45.8\beta_e > \sum_{i=0}^{6} (I'-I^1)[i]= 45.4\beta_e$, the new $I'$ becomes
$I' = I'-I^1 = I_{\{2\}}$.
 \item Create $I^2 = \{4 \beta_e \}$. Since $\sum_{i=0}^{6} I'[i] =
45.4\beta_e \leq \sum_{i=0}^{6} (I'-I^2)[i]= 45.8\beta_e$, $I = I'=
I_{\{2\}}'$ and $\sum_{i=0}^{6} I[i] = 45.4\beta_e$.
\end{enumerate}
\end{exmp}

\section{Analysis}
\label{sec:4}

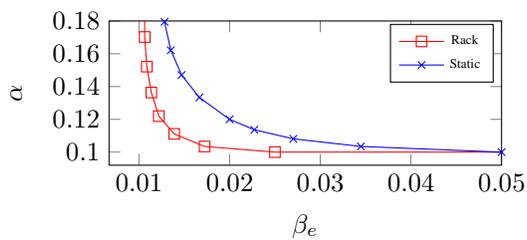
\begin{figure}
\centering

\begin{tikzpicture}
 \begin{axis}
[ xlabel=$\beta_e$,
  ylabel=$\alpha$,
  xmax= 0.05,
  ymax=0.18,
  height=3.5cm,
  width=6.8cm,
  scaled x ticks = false,
  legend style={font=\tiny},
  x tick label style={/pgf/number format/fixed, /pgf/number format/1000 sep =
\thinspace}
]

\addplot[color=red, mark= square] coordinates {(1.1,0.1)
(0.02500000000,0.1000000000)
(0.01724137931,0.1034482759)
(0.01388888889,0.1111111111) (0.01219512195,0.1219512195)
(0.01136363636,0.1363636364) (0.01086956522,0.1521739130)
(0.01063829787,0.1702127660) (0.01063829787, 0.4)};

\addplot[color=blue,mark=x] coordinates { (1.1, 0.1)
(0.05000000000,0.1000000000) (0.03448275862,0.1034482759)
(0.02702702703,0.1081081081) (0.02272727273,0.1136363636)
(0.02000000000,0.1200000000) (0.01666666667,0.1333333333)
(0.01470588235,0.1470588235) (0.01351351351,0.1621621622)
(0.01282051282,0.1794871795) (0.01250000000,0.2000000000)
(0.01250000000, 0.3)};
\legend{Rack, Static}
\end{axis}
\end{tikzpicture}

\caption{Chart comparing the rack model
 with the static cost model for
$M=1$, $k=10$, $d_c^1=5$, $d_c^2=6$, $d=11$, $n_1=n_2=6$ and $\tau=2$. }
 \label{plots:02}
\end{figure}

When $\tau=1$, we have that $\beta_e = \beta_c$, so $\gamma^j=\gamma = d
\beta_e$ for any $j$. This corresponds to the case when the three models
mentioned in this paper coincide in terms of the threshold function,
since we can assume that $\beta_c=\beta_e = \beta$. When $\tau > 1$ and $k \le
d_c^1+1$, the rack model coincides with the static cost model described in
Subsection \ref{subsec:02}.

In order to compare the rack model with the static cost model
when $\tau > 1$ and $k>d_c^1+1$, it is enough to consider the case $r=2$.
Moreover, it only makes sense to consider the equation $C_T^1 = \beta_e (C_c
d_c^1 \tau + C_e d_e^1)$. Using the definitions given for the static cost model and
the rack model, note that $d_c= d_c^1$ and $d_e= d_e^1$. When comparing both
models using $C_T^1$, all the parameters are the same except for $\beta_e= f(i)
= \frac{M}{L[i](k-i)+g(i)}$. Now, we are going to prove that the resulting $L$
will always be greater in the rack model, so both $\beta_e$ and $C_T^1$ will be
less.

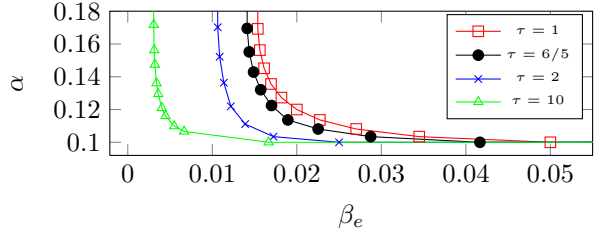
\begin{figure}
\centering
\begin{tikzpicture}
 \begin{axis}
[ xlabel=$\beta_e$,
  ylabel=$\alpha$,
  xmax= 0.055,
  ymax=0.18,
  height=3.5cm,
  width=8cm,
  scaled x ticks = false,
  legend style={font=\tiny},
  x tick label style={/pgf/number format/fixed, /pgf/number format/1000 sep =
\thinspace}
]

\addplot[color=red,mark=square] coordinates { (1.1,
0.1)(0.05000000000,0.1000000000)
(0.03448275862,0.1034482759) (0.02702702703,0.1081081081)
(0.02272727273,0.1136363636) (0.02000000000,0.1200000000)
(0.01818181818,0.1272727273) (0.01694915254,0.1355932203)
(0.01612903226,0.1451612903) (0.01562500000,0.1562500000)
(0.01538461538,0.1692307692) (0.01538461538, 0.3)};

\addplot[color=black,mark=*] coordinates { (1.1,0.1)
(0.04166666667,0.1000000000) (0.02873563218,0.1034482759)
(0.02252252252,0.1081081081) (0.01893939394,0.1136363636)
(0.01700680272,0.1224489796) (0.01572327044,0.1320754717)
(0.01488095238,0.1428571429)(0.01436781609,0.1551724138)
(0.01412429379,0.1694915254) (0.01412429379,0.4)};

\addplot[color=blue,mark=x] coordinates { (1.1,0.1) (0.02500000000,0.1000000000)
(0.01724137931,0.1034482759)
(0.01388888889,0.1111111111) (0.01219512195,0.1219512195)
(0.01136363636,0.1363636364) (0.01086956522,0.1521739130)
(0.01063829787,0.1702127660) (0.01063829787, 0.4)};

\addplot[color=green,mark=triangle] coordinates { (1.1, 0.1)
(0.01666666667,0.1000000000) (0.006666666667,0.1066666667)
(0.005494505494,0.1098901099) (0.004464285714,0.1160714286)
(0.004032258065,0.1209677419) (0.003597122302,0.1294964029)
(0.003401360544,0.1360544218) (0.003205128205,0.1474358974)
(0.003125000000,0.1562500000) (0.003067484663,0.1717791411)
(0.003067484663,0.4)};

\legend{$\tau=1$, $\tau=6/5$,$\tau=2$, $\tau=10$}
\end{axis}
\end{tikzpicture}
\caption{Chart showing the tradeoff curves between $\alpha$ and $\beta_e$ for
$M=1$, $k=10$, $d_c^1=5$,
$d_c^2=6$, $d=11$ and $n_1=n_2=6$, so with $k>d_c^1+1$.}
\label{plots:03}
\end{figure}

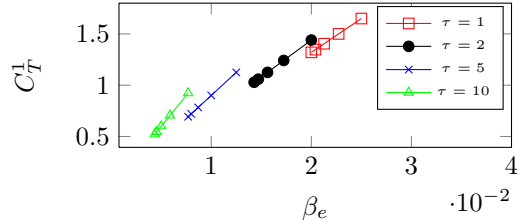
\begin{figure}
\centering

\begin{tikzpicture}
 \begin{axis}
[ xlabel=$\beta_e$,
  ylabel=$C_T^1$,
  xmax= 0.04,
  ymax=1.8,
  height=3.5cm,
  width=6.8cm,
  legend style={font=\tiny},
  x tick label style={/pgf/number format/fixed, /pgf/number format/1000 sep =
\thinspace}
]

\addplot[color=red, mark=square] coordinates {
(0.02500000000,1.650000000)
(0.02272727273,1.500000000) (0.02127659574,1.404255319)
(0.02040816327,1.346938776) (0.02000000000,1.320000000)};

\addplot[color=black,mark=*] coordinates {
(0.02000000000,1.440000000)
(0.01724137931,1.241379310) (0.01562500000,1.125000000)
(0.01470588235,1.058823529) (0.01428571429,1.028571429) };

\addplot[color=blue,mark=x] coordinates {
(0.01250000000,1.125000000)
(0.01000000000,0.9000000000) (0.008695652174,0.7826086957)
(0.008000000000,0.7200000000) (0.007692307692,0.6923076923)};

\addplot[color=green,mark=triangle] coordinates {
 (0.007692307692,0.9230769231)
(0.005882352941,0.7058823529) (0.005000000000,0.6000000000)
(0.004545454545,0.5454545455) (0.004347826087,0.5217391304)};

\legend{$\tau=1$, $\tau=2$, $\tau=5$, $\tau=10$}
\end{axis}
\end{tikzpicture}

\caption{Chart showing the repair cost in
the rack model for $M=1$, $k=5$, $d_c^1=5$, $d_c^2=6$, $d=11$, $n_1=n_2=6$, $C_c
= 1$ and $C_e=10$. The points correspond to the $k=5$ values given by
$f(i)$, $i=0,\ldots,4$. }
 \label{plots:04}
\end{figure}

Assume that the incomes are in terms of $I$. For the static cost model, $I=\{
((d_c^1-i) \tau  +d_e^1) \beta_e \;|\;
i=0,\ldots, d_c^1 \} \cup \{ (d_e^1-i) \beta_e \;|\; i=1,\ldots, k-d_c^1-1 \}$.
Note that $\{(d_e^1-i) \beta_e \;|\; i=1,\ldots, k-d_c^1-1\} = \{(d_c^2-i)
\beta_e \;|\; i=0,\ldots, k-d_c^1-2\}$.
In this case, both models are equal for the first $d_c^1+1$ newcomers,
and different for the remaining $k-d_c^1-1$ newcomers. If $I=I_1
\cup I_3$ for the rack model, the incomes of the remaining $k-d_c^1-1$ newcomers
from the second rack are $(d_c^2-i) \tau \beta_e$, which are greater than
$(d_c^2-i) \beta_e$ of the static cost model. If $I=I_1 \cup I_2$, it can also
be seen that $d_e^1 \beta_e > (d_e^1-i) \beta_e$.
Finally, we can say that the repair cost in the rack model is less than the
repair cost in the static cost model. The comparison between both models is
shown in Figure \ref{plots:02} for an specific example.
The decreasing behavior of $\beta_e$ as $\tau$ increases is shown in Figure
\ref{plots:03} by giving several tradeoff curves for different values of $\tau$.
In Figure \ref{plots:04}, we show that the repair cost is determined by
$\beta_e$, both are directly proportional.

\section{Conclusions}
\label{sec:5}

In this paper, a new mathematical model for a distributed storage environment
where the storage nodes are placed in racks is
presented and analyzed. In this new model, the cost of downloading data units
from nodes in different racks is introduced. That is, the cost of downloading
data units from nodes located in the same rack is much lower than the cost of
downloading data units from nodes located in a different rack. The rack model
is an approach to a more realistic distributed storage environment like the ones
used in companies dedicated to the task of storing information over a network.

Firstly, the rack model is deeply analyzed in the case that there are two racks. The
differences between this model and previous models are shown. Due to it is a
less simplified model compared to the ones presented previously, the rack model
introduces more difficulties in order to be analyzed. The main contribution in this case
is the generalization of the process to find the
threshold function of a distributed storage system. This new generalized threshold function
fits in the previous models and allows to represent the information flow graphs considering
different repair costs. We also provide the tradeoff curve between the repair bandwidth and
the amount of stored data per node and compare it with the ones found in previous
models. We analyze the repair cost of this new model, and we  conclude that
the rack model outperforms previous models in terms of repair cost.

Finally, in this paper, we also study the general rack model where there
are $r \ge 2$ racks. This generalization represents two main contributions: the
modelation of a distributed storage system using any number of racks, and the description of
the algorithm to find the minimum mincut set of newcomers (which is a new
problem compared to the previous models). Once the minimum mincut set is found,
we can apply the same found generalized threshold function for two racks, which is
used to minimize the amount of stored data per node and the
repair bandwidth needed to regenerate a failed node.

It is for further research the case where there are three different costs: one
for nodes within the same rack, another for nodes within different racks but in the
same data center, and a third one for nodes within different data centers. It would be
also important to give some constructions that achieve the optimal bounds. Finally, it is also
interesting to study the possible locality of codes within a rack.

\section*{Acknowledgment}
We would like to thank professor Alexandros G. Dimakis for suggesting us to focus on studying the rack model.  

\bibliographystyle{IEEEtran}
\bibliography{IEEEabrv,references}

\end{document}